\documentclass[a4paper,11pt]{article}
\pdfoutput=1 

\usepackage{jheppub} 

\usepackage[T1]{fontenc} 
\usepackage{amssymb}
\usepackage{amsmath}
\usepackage{color}
\usepackage{xspace}
\usepackage{appendix}
\usepackage{float}
\usepackage{hyperref}
\usepackage{diagbox}
\usepackage[utf8x]{inputenc}
\usepackage{soul}
\usepackage[compat=1.1.0]{tikz-feynhand}

\newcommand{\nn}{\nonumber}
\newcommand{\bea}{\begin{align}}
\newcommand{\eea}{\end{align}}
\newcommand{\beq}{\begin{equation}}
\newcommand{\eeq}{\end{equation}}
\newcommand{\bqa}{\begin{eqnarray}}
\newcommand{\eqa}{\end{eqnarray}}

\newcommand{\mO}{\mathcal{O}}
\newcommand{\mM}{\mathcal{M}}
\newcommand{\ep}{\epsilon}

\def\ksl{k\!\!\!\slash}
\def\psl{p\!\!\!\slash}

\def\epsl{\epsilon\!\!\!\slash}
\allowdisplaybreaks[4]

\title{ One-loop squared  amplitudes for  hadronic $tW$ production at next-to-next-to-leading order in QCD}
\author[a]{Long-Bin Chen,}
\author[b]{Liang Dong,}
\author[b,1]{Hai Tao Li \note{Corresponding author.},}
\author[c,d,e]{Zhao Li,}
\author[b,2]{Jian Wang\note{Corresponding author.},}
\author[b]{and Yefan Wang}

\affiliation[a]{School of physics and materials science, Guangzhou University, Guangzhou 510006, China}
\affiliation[b]{School of Physics, Shandong University, Jinan, Shandong 250100, China}
\affiliation[c]{Institute of High Energy Physics, Chinese Academy of Sciences, Beijing 100049, China}
\affiliation[d]{School of Physics Sciences, University of Chinese Academy of Sciences, Beijing 100039, China}
\affiliation[e]{Center of High Energy Physics, Peking University, Beijing 100871, China}

\abstract{
We present the analytic results of one-loop squared amplitudes for $tW$ production at a hadron collider.
The calculation is performed using the method of differential equations. After renormalization, we have checked that the infrared divergences agree with the general structure predicted by anomalous dimensions. 
The finite remainder contributes to the next-to-next-to-leading order hard function, one of the essential ingredients in the factorization formula of the cross section near the infrared region,
which can be used in resummation of all-order soft gluon effects or a differential next-to-next-to-leading order calculation based on the phase space slicing method.}

\emailAdd{chenlb@gzhu.edu.cn}
\emailAdd{liang.dong@mail.sdu.edu.cn}
\emailAdd{haitao.li@sdu.edu.cn}
\emailAdd{zhaoli@ihep.ac.cn}
\emailAdd{j.wang@sdu.edu.cn}
\emailAdd{wangyefan@sdu.edu.cn}

\begin{document}

\maketitle
\flushbottom

\section{Introduction}
The top quark is the heaviest elementary particle in the Standard Model (SM) and may play an essential role in the electroweak symmetry breaking due to its large coupling with the Higgs boson.
Its short lifetime also allows the study on its various properties as a     ``bare'' quark without hadronization. 
Meanwhile the $W$ boson couples with many SM particles via gauge interactions 
and therefore could be sensitive to the new physics beyond the SM with  such interactions modified.
The recent precise measurement of the CDF collaboration on the $W$ boson mass \cite{CDF:2022hxs}
has shed light on a significant tension with the other experimental observations 
in the assumption of the SM electroweak framework. 
Especially the strong correlation between the top quark and the $W$ boson implies 
that the precise theoretical investigations are indispensable.

\begin{figure}[ht]
\begin{center}
	\begin{minipage}{0.3\linewidth}
		\centering
		\includegraphics[width=0.8\linewidth]{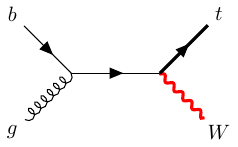}
	\end{minipage}
	\begin{minipage}{0.3\linewidth}
		\centering
		\includegraphics[width=0.8\linewidth]{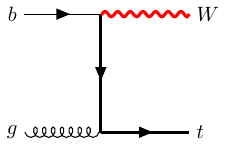}
	\end{minipage}
	\caption{Leading order Feynman diagrams for $gb \to tW$.}
\label{fig:LO}
\end{center}
\end{figure}

The top quark production associated with a $W$ boson, i.e. $tW$-channel, 
as shown in the leading order (LO) Feynman diagrams in figure \ref{fig:LO},
can be used to probe the coupling between the $W$ boson and the top quark, including the Cabibbo-Kobayashi-Maskawa matrix element $V_{tb}$. 
The ATLAS and CMS collaborations at the CERN large hadron collider (LHC) have 
already observed the signal of this process based on the data accumulated 
at center-of-mass energies of $7,8,13$ TeV \cite{Aad:2012xca,Aad:2015eto,Aaboud:2016lpj,Aaboud:2017qyi,ATLAS:2020cwj,Chatrchyan:2012zca,Chatrchyan:2014tua,Sirunyan:2018lcp,CMS:2021vqm}.
So far, only the data recorded in 2016, corresponding to 35.9 ${\rm fb}^{-1}$ integrated luminosity, have been taken into account in the public analyses,
and the uncertainty of the measured cross section is about $11\%$ \cite{Rodriguez:2021ckm}. 
It is expected that the measured cross section of this process will become more precise as more data are going to be analysed in the future. 

To match the accuracy of the experimental measurements, the theoretical predictions must include higher-order radiative effects 
in the perturbative theories.
Since the strong coupling is the largest among the three interactions relevant for fundamental particles,
the QCD corrections are the dominant contribution.
The next-to-leading order (NLO) corrections 
have been obtained  for stable $tW$ final state \cite{Giele:1995kr,Zhu:2001hw,Cao:2008af, Kant:2014oha}
as well as the process including the decay of the top quark and the $W$ boson \cite{Campbell:2005bb}.
The approximate next-to-next-to-next-to-leading order total cross section has been derived by expanding the threshold resummation formula 
\cite{Kidonakis:2006bu,Kidonakis:2010ux,Kidonakis:2016sjf,Kidonakis:2021vob}.
The all-order corrections induced by soft gluons are considered in \cite{Li:2019dhg}.
The effect of parton shower  has been studied in \cite{Frixione:2008yi,Re:2010bp,Jezo:2016ujg}.

However, the complete next-to-next-to-leading order (NNLO) QCD corrections have not been obtained yet,
though a small fraction of the two-loop master integrals have recently been computed analytically \cite{Chen:2021gjv,Long:2021vse}. 
To obtain these corrections, the double-real, real-virtual and double-virtual corrections should be calculated independently and collected together in the end to provide an infrared (IR) finite prediction.
In arranging the cancellation of IR divergences, 
the phase space slicing method \cite{Catani:2007vq} is one of the promising approaches.
Above the slicing cut, which can be the transverse momentum of 
the $tW$ system or the N-jettiness variable \cite{Stewart:2010tn} 
in this process, only the NLO divergence appears 
and it is appropriate to use 
some automatic packages to perform the calculation.
In this part, it is mandatory to consider the interference with the top quark pair production with a top quark decay $\bar{t}\to W\bar{b}$ in order to provide a well-defined perturbative prediction.
Different schemes have been proposed to tackle this issue \cite{Demartin:2016axk}.
Below the slicing cut, a factorization of the cross section 
to several simpler functions holds at the leading power 
in the ratio of the cut over the center-of-mass energy.
Expanding the factorization formula to certain orders 
in the strong coupling gives the required fixed-order contributions.
As one of the ingredients in the factorization formula, 
the N-jettiness soft function has been calculated 
at NNLO by two of the authors \cite{Li:2016tvb,Li:2018tsq}.
The missing part is the NNLO hard function, 
which demands the calculation of one-loop squared 
amplitudes and the interference between two-loop 
and tree-level amplitudes.
In this paper, we present the result of the former.

This paper is organized as follows.
 In section \ref{sec:basis}, we show the kinematics of $tW$ production and the notations used in the following sections. 
 Then we describe the details in calculating the one-loop squared amplitudes in  section \ref{sec:oneloop},
 including the $\gamma_5$ scheme, renormalization procedure, 
 and the infrared divergences cancellation.
 In section \ref{sec:mWexp} we discuss the approximation method by expansion in $m_W^2$.
 Conclusions are made in section \ref{sec:conclusion}.

\section{Kinematics and notations}
\label{sec:basis}

The process $g(k_1)b(k_2)\to W(k_3)t(k_4)$ contains two massive final states with different masses.
For the external particles, there are on-shell conditions $k_1^2=0,k_2^2=0,k_3^2=m_W^2$ and $k_4^2=(k_1+k_2-k_3)^2=m_t^2$.
The Mandelstam variables are defined as
\beq
s=(k_1+k_2)^2\,, \qquad t=(k_1-k_3)^2\,, \qquad u=(k_2-k_3)^2 \,
\eeq
with $s+t+u=m_W^2+m_t^2$.

The LO matrix element is 
\beq \label{eq:mlo}
  \mathcal{M}^{(0)}  = \frac{e\  g_s \ t_{4,2}^{a} }{\sqrt{2} \sin\theta_W}
  \left(
  \frac{\bar{u}(k_4)\epsl_3^* P_L (\ksl_3+\ksl_4)\epsl_1 u(k_2)}{s}
  +
    \frac{\bar{u}(k_4)\epsl_1 (\ksl_2-\ksl_3+m_t)\epsl_3^* P_L u(k_2) }
  { u-m_t^2} 
  \right),
\eeq 
where $t_{4,2}^{a}$ is the generator of the color SU(3) gauge group inserted between the top quark and the bottom quark.
$\epsilon_1^\mu$ and $\epsilon_3^{ *\mu}$ are the polarization vectors for the gluon and $W$ boson, respectively.
$P_L$ is the projection operator to the left-handed quarks.

The polarization summation over all physical polarization states 
for the $W$ boson and the gluon is 
\begin{align}
\sum_i \epsilon_i^{*\mu}(k_3)\epsilon_i^{\nu}(k_3) = -g^{\mu\nu} + \frac{k_3^{\mu} k_3^{\nu}}{m_W^2}
\label{eq:pol-sum}
\end{align} 
and
\begin{align} 
\sum_i \epsilon_i^{\mu}(k_1)\epsilon_i^{*\nu}(k_1) = -g^{\mu\nu} + \frac{k_1^{\mu} n^{\nu} + k_1^{\nu} n^{\mu}}{k_1\cdot n},
\end{align}
respectively. Here $n$ is a light-like four-vector.
Consequently, the ghost field is not needed in the external state
in the loop amplitudes.
In practice,  we can simply use 
\begin{align}
\sum_i \epsilon_i^{\mu}(k_1)\epsilon_i^{*\nu}(k_1) = -g^{\mu\nu}.
\label{eq-pol-sum}
\end{align}
due to  the Ward identity\footnote{In order to 
check the Ward identity, all the external particles 
are required to be physical. 
If there are more than two gluons in the external state, 
e.g., $gb\to Wtg$, 
it is not appropriate to 
use eq.(\ref{eq-pol-sum}) for gluons.}.
As a result, we obtain the LO amplitude squared
\begin{align}
   \overline{ |\mathcal{M}^{(0)}|^2} =& \frac{e^2g_s^2 C_F}{32 \sin^2\theta_W~s~(u-m_t^2)^2~ m_W^2}\Bigl[m_t^8-m_t^6(2s+u)+m_t^4\bigl(-2m_W^4-2m_W^2u+(s+u)^2\bigr)\notag\\
    &+m_t^2\bigl(4m_W^6-2m_W^4u+2m_W^2(s^2-su+2u^2)-u(s+u)^2\bigr)\notag\\
    &-2m_W^2u\bigl(2m_W^4-2m_W^2(s+u)+s^2+u^2\bigr)\Bigr],
\end{align}
where the factors from spin and color average 
in the initial states have been included.

The LO  hadronic cross section for $tW$ production  at the LHC is 
\begin{align}
    \sigma = \int_0^1 dx_1 dx_2 \frac{1}{2s}\int d\Phi_2  \left[ f_{g/p}(x_1,\mu_f) f_{b/p}(x_2,\mu_f)\overline{|\mathcal{M}^{(0)}|^2} +(x_1\leftrightarrow x_2)\right],
\end{align}
where $d\Phi_2$ is the Lorentz invariant two-body phase space,
and $f_{i/p}(x,\mu_f)$ is the parton distribution function at the factorization scale $\mu_f$.

In order to obtain the complete NNLO QCD corrections to this process, 
we need to obtain 
one-loop virtual corrections to $\mM_{gb\to Wt}$ up to $\mO(\ep^2)$,
two-loop virtual corrections to $\mM_{gb\to Wt}$ up to $\mO(\ep^0)$,
one-loop virtual corrections to $\mM_{gb\to Wt+j}$ up to $\mO(\ep^0)$
\footnote{It is not necessary to compute this part up to $\mO(\ep^2)$ because
the scattering amplitude reduces to $\mM_{gb\to Wt}$ 
when the integration of three-body phase space develops a pole in $\ep$.},
and tree-level $\mM_{gb\to Wt+jj}$.
In this paper, we focus on the first contribution, 
leaving the others and the combination of all contributions after phase space integration
to the future work.

\section{One-loop calculation}
\label{sec:oneloop}

\subsection{Bare loop amplitudes}

We generate all the one-loop virtual diagrams as well as the diagrams with counterterms by FeynArts \cite{Hahn:2000kx}. 
The corresponding amplitudes can be written down automatically and manipulated by FeynCalc \cite{Shtabovenko:2020gxv}.
We do not keep the polarization information at the moment and 
focus only on amplitude squared, i.e.,
we calculate the interference between the one-loop diagrams 
and tree-level or one-loop diagrams.
All the Lorentz indices are contracted after spin sums, 
and therefore the loop integrals appear in the form of scalar integrals,
though there may be some loop momenta in the numerators.
The spin and polarization information may be useful if one wants to study the kinematic distribution of the decay products of the top quark and the $W$ boson.
This can be obtained by using the spinor-helicity formalism for the external states after decay,
as in \cite{Campbell:2005bb}
for the NLO corrections, or by replacing the polarization vector of specific helicities by a linear combination of external momenta \cite{Chen:2019wyb}.
Our calculation can be readily extended 
to the helicity amplitudes because they share the same master integrals as the spin-summed amplitudes.

We have chosen the anticommuting $\gamma_5$ scheme following ref.~\cite{Korner:1991sx} to calculate the traces in the squared amplitude.
In this scheme, it is easy to deal with the traces with two (or more generally even number of) $\gamma_5$ matrices in our case due to the anticommutativity property.
In the trace with a single (or more generally odd number of) $\gamma_5$, 
since the cyclicity in traces has been sacrificed,
we choose the $Wtb$ vertex in the amplitude 
as a {\it reading point}, i.e. the starting vertex in a trace.
The non-vanishing trace with one $\gamma_5$ 
would contribute an anti-symmetric $\varepsilon_{\mu\nu\rho\sigma}$ tensor 
to each term in the result.
Its Lorentz indices should contract only with those of the external momenta
in the spin summed amplitude squared, given that there exists only one fermion line.
Since only three independent momenta are involved in $tW$ production process,
the traces with one $\gamma_5$ would vanish at the end.

Then we use two different methods to reduce all the scalar integrals 
in the squared amplitude 
to a set of basis integrals, called master integrals.
In the first method, 
the reduction is performed following the Passarino-Veltman procedure \cite{Passarino:1978jh},
and the master integrals can be found from the literature \cite{Ellis:2007qk}
or calculated by Package-X \cite{Patel:2015tea}
up to $O(\epsilon^0)$.
They can also be evaluated up to higher orders in $\ep$ using the AMFlow package \cite{Liu:2022chg}.
In the other method, we use integration-by-parts (IBP) technique to 
derive relations among the scalar integrals and carry out the
reduction using FIRE \cite{Lee:2012cn}. The master integrals are calculated using the 
method of differential equations \cite{Kotikov:1990kg,Kotikov:1991pm}. 
After transforming the differential equations to a canonical form \cite{Henn:2013pwa}, the solutions can be readily obtained 
 up to any orders in $\epsilon$.
It has been checked analytically that both methods lead to the same results 
for the interference of one-loop and tree-level amplitudes.
For the one-loop squared amplitudes, the coefficients of $\epsilon^{-4}$ and $\epsilon^{-3} $ 
coincide in analytical form,
while the coefficients of  $\epsilon^{-2},\ep^{-1}$ and $\ep^0$  agree numerically. 
We also notice that there are more master integrals 
using Passarino-Veltman reduction than IBP relations.
The coefficients of the master integrals in the former contain no divergences, while those in the latter do.
These differences indicate that the agreement between the two methods is highly non-trivial.
In the following, we present more details about the IBP method.
 
We find that
 all the master integrals in the one-loop diagrams could be expressed in terms of three families of master integrals as: 
\bqa
I^1_{n_1,n_2,n_3,n_4} & = & \int{\mathcal D}^D q_1~\frac{1}{D_1^{n_1}~D_2^{n_2}~D_3^{n_3}~D_4^{n_4}},\nonumber \\
I^2_{n_1,n_2,n_3,n_4} & = & \int{\mathcal D}^D q_1~\frac{1}{D_1^{n_1}~D_2^{n_2}~D_5^{n_3}~D_4^{n_4}},\nonumber\\
I^3_{n_1,n_2,n_3,n_4} & = & \int{\mathcal D}^D q_1~\frac{1}{D_1^{n_1}~D_6^{n_2}~D_3^{n_3}~D_4^{n_4}}\nonumber
\label{def}
\eqa
with
\beq
{\mathcal D}^D q_1 = \frac{\left(m_t^2\right)^\epsilon}{i \pi^{D/2}\Gamma(1+\epsilon)}  d^D q_1
\eeq
and
\begin{align}
    D_1 & = -q_1^2\;,  \nn 
    \\ D_2&= -(q_1+k_2)^2\;,  \nn 
    \\ D_3&= -(q_1+k_1+k_2)^2\;, \nn 
    \\ D_4 & = -(q_1+k_4)^2+m_t^2\;,  \nn 
    \\ D_5 &= -(q_1+k_2-k_3)^2+m_t^2\;,  \nn 
    \\ D_6&=-(q_1+k_1)^2\,.
\end{align}
We will present the details for the calculation of the first family
in the text, but leave the second family to the appendix \ref{sec:diffB}.
The third family can be related to the first using $k_1\leftrightarrow k_2$,
and thus we will not present the results for the master integrals in it.

For the squared one-loop amplitude,
the reduction of any general integral to master integrals can be carried out
as the traditional two-loop integrals.
It is interesting to notice that additional Lorentz invariant combinations, such as  $(q_1\cdot q_2)^3$ with $q_2$ being another 
loop momentum, can appear in the squared amplitude after summing the polarizations of the vector bosons. 
But they do not appear in the master integrals.
As a consequence, we need the product of one integral
in any of the above families and another in complex conjugate form.

The canonical basis of the first family is chosen to be
\bqa
M_1&=&\epsilon\, I^1_{0,0,0,2}~,\nonumber\\
M_2&=&\epsilon\, t I^1_{0,1,0,2}~,\nonumber\\
M_3&=&\epsilon\, m_W^2 I^1_{0,0,1,2}~,\nonumber\\
M_4&=&\epsilon\, s I^1_{1,0,2,0}~,\nonumber\\
M_5&=&\epsilon^2\, \sqrt{(s-(m_W-m_t)^2)}\sqrt{(s-(m_W+m_t)^2)} I^1_{1,0,1,1}~,\nonumber\\
M_6&=&\epsilon^2\, s(t-m_t^2)I^1_{1,1,1,1}~.
\label{eq:MIs}
\eqa
To rationalize the roots, we define the variables $x,y,z$ via
\bqa
s &=& m_t^2\, \frac{(x+z)(1+x z)}{x}, \nonumber \\
t &=& y\, m_t^2, \nonumber \\
m_W &=& z\, m_t.
\label{xyzd}
\eqa
The master integrals in eq.(\ref{eq:MIs}) satisfy the differential equation
\bqa
d\, \text{{\bf M}}(x,y,z;\epsilon)=\epsilon\, (d \, \tilde{A})\,  \text{{\bf M}}(x,y,z;\epsilon)
\eqa
with
\bqa
d\, \tilde{A}=\sum_{i=1}^{12} R_i\,  d \ln(l_i)~.
\eqa
The arguments $l_i$ of this $d$\,ln form encode all the
dependence of the differential equations on the kinematics, are called the {\it alphabet}, consisting of the following letters,
\begin{align}
\begin{alignedat}{2}
l_1 & =x\,,&\quad
l_2 & =x+1\,, \\
l_3 & =x-1\,, &\quad
l_4 & =x+z\,, \\
l_5 & =x\,z+1\,,&\quad
l_6 & =x~y+z\,,\\
l_7 & =x\,z+y  \,,&\quad
l_8 & =y \, , \\
l_9 & =y-1\,,&\quad
l_{10} & =y-z^2\, , \\
l_{11} & =z\,,&\quad
l_{12} & =z^2-1\,.
\end{alignedat} \stepcounter{equation}\tag{\theequation}
\label{alphabet}
\end{align}
$R_i$ are rational matrices, given by
\bqa
R_1&=&\left(
\begin{array}{cccccc}
 0 & 0 & 0 & 0 & 0 & 0 \\
 0 & 0 & 0 & 0 & 0 & 0 \\
 0 & 0 & 0 & 0 & 0 & 0 \\
 0 & 0 & 0 & 1 & 0 & 0 \\
 1 & 0 & 0 & -1 & 0 & 0 \\
 1 & 2 & 0 & 0 & 0 & 0 \\
\end{array}\right),~~
R_2=\left(
\begin{array}{cccccc}
 0 & 0 & 0 & 0 & 0 & 0 \\
 0 & 0 & 0 & 0 & 0 & 0 \\
 0 & 0 & 0 & 0 & 0 & 0 \\
 0 & 0 & 0 & 0 & 0 & 0 \\
 0 & 0 & 0 & 0 & 2 & 0 \\
 0 & 0 & 0 & 0 & 0 & 0 \\
\end{array}
\right),~~
R_3=\left(
\begin{array}{cccccc}
 0 & 0 & 0 & 0 & 0 & 0 \\
 0 & 0 & 0 & 0 & 0 & 0 \\
 0 & 0 & 0 & 0 & 0 & 0 \\
 0 & 0 & 0 & 0 & 0 & 0 \\
 0 & 0 & 0 & 0 & 2 & 0 \\
 0 & 0 & 0 & 0 & 0 & 0 \\
\end{array}
\right), \nonumber\\
R_4&=&\left(
\begin{array}{cccccc}
 0 & 0 & 0 & 0 & 0 & 0 \\
 0 & 0 & 0 & 0 & 0 & 0 \\
 0 & 0 & 0 & 0 & 0 & 0 \\
 0 & 0 & 0 & -1 & 0 & 0 \\
 -1 & 0 & -2 & 1 & -2 & 0 \\
 -\frac{1}{2} & 0 & -1 & 1 & -1 & -1 \\
\end{array}
\right),~~
R_5=\left(
\begin{array}{cccccc}
 0 & 0 & 0 & 0 & 0 & 0 \\
 0 & 0 & 0 & 0 & 0 & 0 \\
 0 & 0 & 0 & 0 & 0 & 0 \\
 0 & 0 & 0 & -1 & 0 & 0 \\
 1 & 0 & 2 & -1 & -2 & 0 \\
 -\frac{1}{2} & 0 & -1 & 1 & 1 & -1 \\
\end{array}
\right),\nonumber\\
R_6 &=& \left(
\begin{array}{cccccc}
 0 & 0 & 0 & 0 & 0 & 0 \\
 0 & 0 & 0 & 0 & 0 & 0 \\
 0 & 0 & 0 & 0 & 0 & 0 \\
 0 & 0 & 0 & 0 & 0 & 0 \\
 0 & 0 & 0 & 0 & 0 & 0 \\
 -\frac{1}{2} & -2 & 1 & -1 & 1 & 1 \\
\end{array}
\right),~~
R_7 =\left(
\begin{array}{cccccc}
 0 & 0 & 0 & 0 & 0 & 0 \\
 0 & 0 & 0 & 0 & 0 & 0 \\
 0 & 0 & 0 & 0 & 0 & 0 \\
 0 & 0 & 0 & 0 & 0 & 0 \\
 0 & 0 & 0 & 0 & 0 & 0 \\
 -\frac{1}{2} & -2 & 1 & -1 & -1 & 1 \\
\end{array}
\right),\nonumber\\
R_8 &=&\left(
\begin{array}{cccccc}
 0 & 0 & 0 & 0 & 0 & 0 \\
 0 & 1 & 0 & 0 & 0 & 0 \\
 0 & 0 & 0 & 0 & 0 & 0 \\
 0 & 0 & 0 & 0 & 0 & 0 \\
 0 & 0 & 0 & 0 & 0 & 0 \\
 0 & 0 & 0 & 0 & 0 & 0 \\
\end{array}
\right),~~
R_9=\left(
\begin{array}{cccccc}
 0 & 0 & 0 & 0 & 0 & 0 \\
 -1 & -2 & 0 & 0 & 0 & 0 \\
 0 & 0 & 0 & 0 & 0 & 0 \\
 0 & 0 & 0 & 0 & 0 & 0 \\
 0 & 0 & 0 & 0 & 0 & 0 \\
 1 & 2 & 0 & 0 & 0 & -2 \\
\end{array}
\right),\nonumber\\
R_{10}&=&\left(
\begin{array}{cccccc}
 0 & 0 & 0 & 0 & 0 & 0 \\
 0 & 0 & 0 & 0 & 0 & 0 \\
 0 & 0 & 0 & 0 & 0 & 0 \\
 0 & 0 & 0 & 0 & 0 & 0 \\
 0 & 0 & 0 & 0 & 0 & 0 \\
 0 & 2 & -2 & 0 & 0 & 0 \\
\end{array}
\right),~~
R_{11}=\left(
\begin{array}{cccccc}
 0 & 0 & 0 & 0 & 0 & 0 \\
 0 & 0 & 0 & 0 & 0 & 0 \\
 0 & 0 & 2 & 0 & 0 & 0 \\
 0 & 0 & 0 & 0 & 0 & 0 \\
 0 & 0 & 0 & 0 & 2 & 0 \\
 0 & 0 & 0 & 0 & 0 & 0 \\
\end{array}
\right),~~
R_{12}=\left(
\begin{array}{cccccc}
 0 & 0 & 0 & 0 & 0 & 0 \\
 0 & 0 & 0 & 0 & 0 & 0 \\
 -1 & 0 & -2 & 0 & 0 & 0 \\
 0 & 0 & 0 & 0 & 0 & 0 \\
 0 & 0 & 0 & 0 & 0 & 0 \\
 1 & 0 & 2 & 0 & 0 & 0 \\
\end{array}
\right).\nonumber
\eqa

 In order to solve the above differential equations, 
 we need to specify the boundary conditions.
The integrals $M_1$ and $M_4$ are simple enough so that
 one can get the analytical results with full dependence on $\epsilon$
 from direct computation.  Since all the one-loop integrals we encounter do not have any branch cut at the kinematic points corresponding to $t=0, m_W=0,s=(m_t+m_W)^2,$ and $u t=m_t^2 m_W^2$, they are regular without any singularity at these kinematic points.
 The boundary condition of $M_2$ is obtained using 
the fact that it is regular at $t=0$. Explicitly,
the coefficient of $1/t$ in the differential equation, which is a linear combination of some master integrals, is vanishing.
Similarly, the boundary conditions of $M_3,M_5$ and $M_6$ are derived from the regular condition at $m_W=0$, $s=(m_t+m_W)^2$ and $u t=m_t^2 m_W^2$, respectively. 

Since the  letters are just polynomials in $x,y,z$,
we can solve the differential equation recursively using
the boundaries discussed above and write 
the final result in terms of the multiple polylogarithms 
\cite{Goncharov:1998kja},
which are defined as $G(x)\equiv 1$ and
\bqa
G_{a_1,a_2,\ldots,a_n}(x) &\equiv & \int_0^x \frac{\text{d} t}{t - a_1} G_{a_2,\ldots,a_n}(t)\, ,\\
G_{\overrightarrow{0}_n}(x) & \equiv & \frac{1}{n!}\ln^n x\, .
\eqa
The dimension of the vector $(a_1,a_2,\ldots,a_n)$ is usually called the transcendental $weight$ of the multiple polylogarithm.
We need multiple polylogarithms of at most transcendental weight four in the one-loop squared amplitudes.

\subsection{Renormalization}

The loop amplitude computed above contains ultra-violate (UV) and IR divergences. 
The UV divergences cancel with the contribution from counter-terms,
arising from the renormalization of the couplings, masses and field strength.

In general, there are two methods to calculate the contribution of counter-terms.
In one method, the bare Lagrangian is used to generate the Feynman rules and  the loop amplitudes,
and the external wave function renormalization is taken into account according to the Lehmann–Symanzik–Zimmermann (LSZ)  reduction formula.
We introduce the renormalization factor $Z_2^g, Z_2^b, Z_2^t$ for the gluon field, bottom quark field and top quark field, respectively, 
by defining the renormalized fields as $\psi_{\rm bare} = Z_2^{1/2} \psi_r, A^{\mu}_{\rm bare} = (Z_2^g)^{1/2} A^{\mu}_r$.
We do not consider the renormalization of the $W$ boson field 
since we focus only  on QCD corrections.
Then the bare masses and couplings in the result should be replaced by the physical masses and couplings.
The relation between the bare top quark mass $m_{t,{\rm bare}}$ and the physical mass $m_{t}$ is $m_{t,{\rm bare}}=Z^t_m m_{t}$.
The strong coupling $\alpha_s$ is renomalized in the $\overline{\rm MS}$ scheme
through the identity $\alpha_s^{\rm bare}C_{\ep} =   \mu^{2\ep} \alpha_s^{\rm ren} Z_{\alpha_s}$
\footnote{For simplicity, we use $\alpha_s$ to denote $\alpha_s^{\rm ren}$ everywhere in the paper. }
with $ C_{\ep} = (4\pi e^{- \gamma_E} )^{\ep} $.
The perturbative expansion of these renormalization factors is listed in the appendix \ref{sec:renormfactor}.
As a result, the UV renormalized scattering amplitude up to two loops is 
\begin{align}
  \mathcal{M}_{\rm ren} &= \mathcal{M}^{(0)}_{\rm bare} + \frac{\alpha_s^{\rm bare}} {4\pi} ( \mathcal{M}^{(1)}_{\rm bare}+ \mathcal{M}^{(1)}_{\rm C.T.})+\left(\frac{\alpha_s^{\rm bare}} {4\pi} \right)^2( \mathcal{M}^{(2)}_{\rm bare}+ \mathcal{M}^{(2)}_{\rm C.T.})
  \nn \\
  &= \mathcal{M}_{\rm ren}^{(0)} +\frac{\alpha_s} {4\pi} \mathcal{M}_{\rm ren}^{(1)} +\left(\frac{\alpha_s} {4\pi} \right)^2 \mathcal{M}_{\rm ren}^{(2)},
  \label{eq:ampren}
\end{align}
where $\mathcal{M}_{\rm bare}$ is the bare  amplitude and  $\mathcal{M}_{\rm C.T.}$ denotes
the amplitude whenever higher-order results of the  renormalization factors are used.
In the first line, we express the results in terms of bare quantities, while the second line contains only renormalized ones. 
There is not a one-to-one correspondence between the two lines, e.g., the third term in the second line could receive contributions from the second term in the first line.

The other method is to use the Lagrangian in terms of renormalized fields and couplings, which includes new counter-term vertices.
The top quark two-point counter-term vertex is described by 
$i[\psl (Z_2^t-1) - (Z_2^t Z_m^t-1) m_t]$.
The combination of the top quark propagator and the two-point counter-terms now becomes 
\begin{align}
    \frac{i (Z_2^t)^{-1}}{\psl-m_t}\left(1 + \frac{m_t \delta Z_m^t}{\psl-m_t}+ \frac{(m_t \delta Z_m^t)^2}{(\psl-m_t)^2}+\cdots \right)\,,
    \label{eq:quarkpropren}
\end{align}
where we have used the notation $\delta Z\equiv Z-1$.
The omitted terms do not contribute to an NNLO calculation.
Similar expressions can be obtained for the massless bottom quark
by just neglecting all the mass terms.
The gluon and quark three-point counter-term vertex is given by 
a multiplicative factor $\delta_1^q = Z_{g_s} Z_2^q (Z_2^g)^{1/2}-1,q=t,b$.
The $Wtb$ counter-term vertex has a factor $\delta_1^W= (Z_2^t)^{1/2} (Z_2^b)^{1/2}-1$. 
At two loops, we would also need the gluon two-point and three-point counter-terms.
Furthermore, no wave function renormalization factors are needed in the LSZ formula because physical fields have been employed. 
Therefore, the renormalized scattering amplitude up to NNLO is given by 
\begin{align}
  \mathcal{M}_{\rm ren} &= \mathcal{M}^{(0)}_{\rm ren} + \frac{\alpha_s} {4\pi} ( \mathcal{M}^{(1)}_{\rm loop}+ \mathcal{M}^{(1)}_{\rm C.T.})+\left(\frac{\alpha_s} {4\pi} \right)^2( \mathcal{M}^{(2)}_{\rm loop}+ \mathcal{M}^{(2)}_{\rm C.T.})
  \nn \\
  &= \mathcal{M}_{\rm ren}^{(0)} +\frac{\alpha_s} {4\pi} \mathcal{M}_{\rm ren}^{(1)} +\left(\frac{\alpha_s} {4\pi} \right)^2 \mathcal{M}_{\rm ren}^{(2)}\,,
  \label{eq:ampren2}
\end{align}
where $\mathcal{M}_{\rm loop}$ is the  loop amplitude  and  $\mathcal{M}_{\rm C.T.}$ denotes
the amplitude with counter-term vertices inserted  in a diagram anywhere 
possible except for the external legs.
The UV divergences cancel between $\mathcal{M}^{(i)}_{\rm loop}$ and $\mathcal{M}^{(i)}_{\rm C.T.}$, and thus there is  a one-to-one correspondence between the two lines in the above equation.

The two renormalization methods are equivalent, which can be understood 
by observing the cancellation of the wave function renormalization factors in eq.(\ref{eq:quarkpropren}) and those in the combination of tree-level and counter-terms vertices\footnote{An analogous  equation to eq.(\ref{eq:quarkpropren}) for the gluon propagator including counter-term vertices can be obtained in the Landau gauge. }.
As a result, only the renormalization factors associated with the external particles, as well as  $\delta Z_{\alpha_s}$ and $\delta Z_{m}^t$, 
should be considered.

In eq.(\ref{eq:ampren2}),
we have used a superscript to denote the order in $\alpha_s$ for the perturbative expansion of the amplitude.
Notice that $\mathcal{M}^{(2)}_{\rm C.T.}$ contains not only the tree-level diagrams with counter-terms at $\alpha_s^2$ but also the one-loop diagrams with counter-terms at $\alpha_s$.
Since the one-loop integrals contain up to $1/\ep^2$ poles,
we need the counter-terms at $\mO(\alpha_s)$ up to $\mO(\ep^2)$ in $\mathcal{M}^{(2)}_{\rm C.T.}$.
They are also needed when we calculate the squared amplitude, e.g.,
$\mathcal{M}^{(1)}_{\rm loop}(\mathcal{M}^{(1)}_{\rm C.T.})^*$.

\subsection{Infrared divergences}

After taking into account the contribution of counter-terms,
the UV divergences disappear.
But there are still IR divergences.
These will cancel with the corresponding contribution of real corrections \footnote{At hadron colliders, renormalization of the parton distribution function is also needed.} in the prediction for a physical observable.
This cancellation should be checked analytically before performing any reasonable numerical computation.
Though the real correction is out of the scope of this paper,
we can make a non-trivial check based on the understanding 
of the factorization property of amplitudes in the IR limit.
More precisely, it has been proven that the UV renormalized amplitude can be written in the form
\begin{align}
    \mathcal{M}_{\rm ren} 
    =\mathbf{Z}\mathcal{M}_{\rm fin} \,,
\end{align}
where all the IR divergences are encoded in the factor  $\mathbf{Z}$ 
and the remainder is finite as $\ep\to 0$ \cite{Becher:2009cu,Becher:2009qa,Becher:2009kw,Ferroglia:2009ep}.
Following our convention, both the $\mathbf{Z}$ factor and the finite part have an expansion in $\alpha_s$,
\begin{align}
    \mathbf{Z}
    & = 1+ \frac{\alpha_s}{4\pi} \mathbf{Z}^{(1)}+\left(\frac{\alpha_s}{4\pi} \right)^2\mathbf{Z}^{(2)}+\cdots\,,\nn \\ 
    \mathcal{M}_{\rm fin} 
    & = \mathcal{M}_{\rm fin}^{(0)} + \frac{\alpha_s}{4\pi} \mathcal{M}_{\rm fin}^{(1)}+\left(\frac{\alpha_s}{4\pi} \right)^2\mathcal{M}_{\rm fin}^{(2)}+\cdots\,    .
\end{align}
Then it is ready to get
\begin{align}
   \mathcal{M}_{\rm fin}^{(0)} &= \mathcal{M}_{\rm ren}^{(0)}\,,
   \nn \\
   \mathcal{M}_{\rm fin}^{(1)} &= \mathcal{M}_{\rm ren}^{(1)}-\mathbf{Z}^{(1)} \mathcal{M}_{\rm ren}^{(0)}\,,
   \nn \\
   \mathcal{M}_{\rm fin}^{(2)} & =  \mathcal{M}_{\rm ren}^{(2)}+ ((\mathbf{Z}^{(1)})^2-\mathbf{Z}^{(2)})\mathcal{M}_{\rm ren}^{(0)}  -\mathbf{Z}^{(1)} \mathcal{M}_{\rm ren}^{(1)} \,.
\end{align}
The factor $\mathbf{Z}$ has been computed  up to two-loop order for a general process with massive colored particles~\cite{Becher:2009kw,Ferroglia:2009ep,Mitov:2010xw},
and to three-loop order for single top production \cite{Kidonakis:2019nqa}.
In the present paper,
we are interested in the one-loop squared amplitudes,
\begin{align}\label{eq:fin}
    \Big| \mathcal{M}^{(1)}_{\rm fin} \Big|^2 =& \Big| \mathcal{M}^{(1)}_{\rm ren} \Big|^2 +  \Big| \mathbf{Z}^{(1)} \mathcal{M}_{\rm ren}^{(0)} \Big|^2- (\mathbf{Z}^{(1)} \mathcal{M}_{\rm ren}^{(0)} \mathcal{M}^{(1)*}_{\rm ren}+\mathbf{Z}^{(1)*} \mathcal{M}_{\rm ren}^{(0)*}\mathcal{M}_{\rm ren}^{(1)}),
\end{align}
where  the  $\mathbf{Z}$ factor at one loop is given by
\begin{align}
   \mathbf{Z}^{(1)} =&  -(C_A+ C_F)\frac{\gamma^{(0)}_{\rm cusp}}{4\epsilon^2} +\frac{\gamma_g^{(0)}+\gamma_b^{(0)}+\gamma_t^{(0)}}{2\epsilon}
    \nn \\ 
   &  +\frac{\gamma_{\rm cusp}^{(0)}}{4 \epsilon} \left(-C_A \ln\frac{\mu^2}{-s} -C_A \ln \frac{\mu m_t }{m_t^2-u} +(C_A-2C_F)\ln\frac{\mu m_t }{m_t^2-t} \right)  .
\end{align}
The anomalous dimensions $\gamma_{\rm cusp}$, $\gamma_g$, $\gamma_b$ and $\gamma_t$ can be found in the appendix of ref.~\cite{Li:2013mia}. 
Notice that we have chosen such a factorization scheme that $\mathbf{Z}$ does not have any finite part, which corresponds to the $\overline{{\rm MS}}$ scheme.
The presence of double poles in $\mathbf{Z}^{(1)} $ 
requires the squared amplitudes $\mathcal{M}_{\rm ren}^{(0)} \mathcal{M}^{(1)*}_{\rm ren}$ and $\mathcal{M}_{\rm ren}^{(1)} \mathcal{M}^{(0)*}_{\rm ren}$ 
to be calculated up to $\mathcal{O}(\ep^2)$
because this could contribute to the $\mathcal{O}(\ep^0)$ 
part in $ \Big| \mathcal{M}^{(1)}_{\rm fin} \Big|^2 $.

\begin{figure}
    \centering
    \includegraphics[width = 0.49 \textwidth]{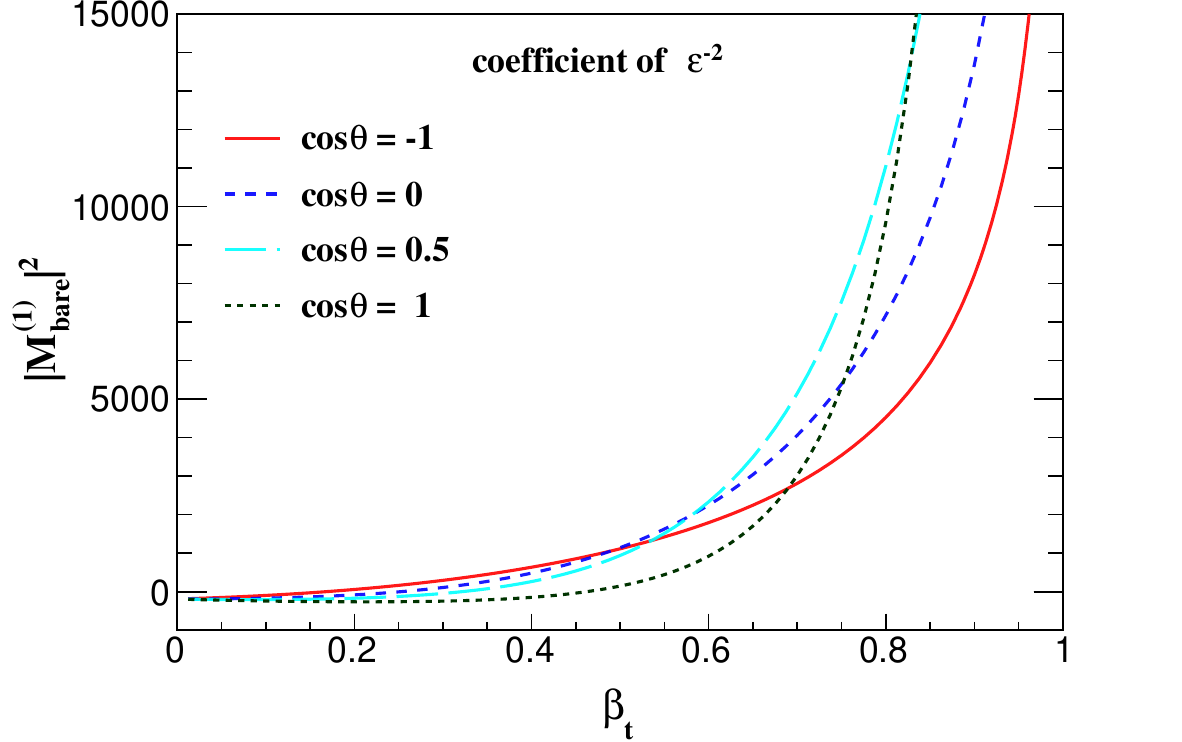}
    \includegraphics[width = 0.49 \textwidth]{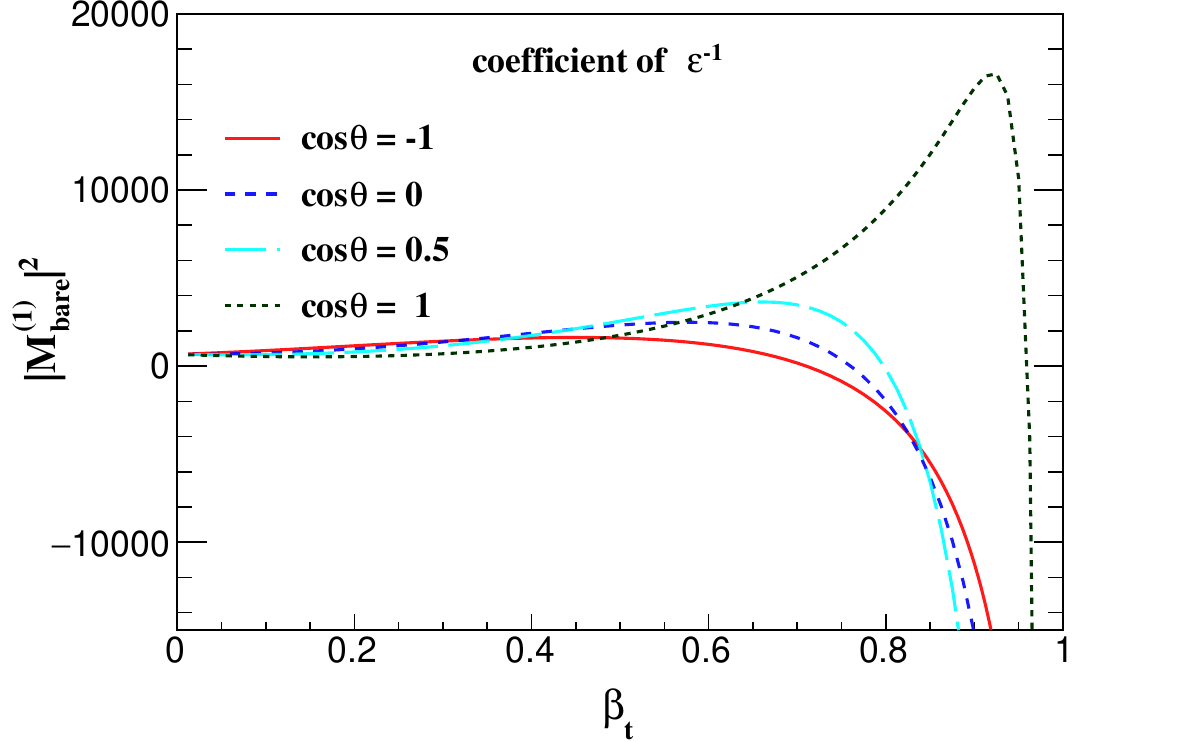}
    \caption{The coefficients of $\ep^{-2}$ (left) and $\ep^{-1}$ (right) in the bare loop squared amplitude as a function of $\beta_t$ with several fixed values of  $\cos\theta$. The red, blue, cyan and dark green lines are for $\cos\theta=-1$, $\cos\theta=0$, $\cos\theta=0.5$ and $\cos\theta=1$, respectively. We fix the mass ratio $m_W/m_t=\sqrt{3/14}$ and extract a global factor $e^2 g_s^2/\sin^2{\theta_W}$. The renormalization scale is set to be $\mu=m_t$.}
    \label{fig:bare}
\end{figure}

The second term on the right hand side of eq.(\ref{eq:fin})
contains $\ep^{-4}$ poles, which should cancel
with those in the first and third terms.
As a result, we can check the coefficient of $\ep^{-4}$ in 
$\Big| \mathcal{M}^{(1)}_{\rm ren} \Big|^2$
and in turn that in $\Big| \mathcal{M}^{(1)}_{\rm bare} \Big|^2$ analytically,
given that the result with $\mathcal{M}^{(1)}_{\rm C.T.}$ is easier to obtain.
Similar cancellation has been checked analytically for the  coefficient of $\ep^{-3}$.
However, it is nontrivial to check the cancellation
of the  coefficients of $\ep^{-2}$ and $\ep^{-1}$  analytically
because the $\mathcal{O}(\ep^0)$ and $\mathcal{O}(\ep^1)$
parts of $\mathcal{M}^{(1)}_{\rm bare}$ would get involved.
Instead, we confine the cancellation to be numerical with high precision.
In figure~\ref{fig:bare}, we present our numerical results for
these coefficients.
In the plots, we have used the parameters $\beta_t$  and $\cos\theta$ that are very suitable to describe
the kinematics of the collision process in a finite range.
More specifically, the top quark momentum is written as $k_4^{\mu} = E_t (1, \beta_t \sin \theta \cos\phi, \beta_t \sin\theta\sin\phi, \beta_t \cos\theta)$ in the center-of-mass frame of the incoming partons,
where $\beta_t=\sqrt{1-m_t^2/E_t^2}$ measures the velocity of the top quark and $\theta$ is the angle between the initial-state gluon and the final-state top quark. 

It can be seen that the magnitudes of the coefficients span over several orders, especially growing or decreasing very fast as $\beta_t\to 1$.
We have checked that the cancellation 
of these coefficients with the other parts in eq.(\ref{eq:fin})
happens at the level of  at least eight digits
over almost the whole range of $\beta_t$ and $\cos\theta$
\footnote{  We do not cover the phase space points with  $\beta_t = 1$, where the coefficients become divergent due to new collinear singularities. But in practice, we will not reach such a point since the top quark mass is not vanishing (or the collider energy is limited).}.
Actually, it is not an easy task to achieve this.
The main obstacle is the precise calculation of the very lengthy coefficients of the master integrals in the squared amplitudes,
which contain high powers of $s$, $t$ and $u$ variables.
And they are represented in terms of $\beta_t$ and $\cos\theta$ as 
\begin{align}
    s&=m_W^2+\frac{1+\beta_t^2}{1-\beta_t^2} m_t^2 + \frac{2 m_t}{1-\beta_t^2}\sqrt{m_W^2+\beta_t^2(m_t^2-m_W^2)}\,,
    \nn \\ 
    t&=m_t^2-\frac{1+\beta_t\cos\theta}{1-\beta_t^2} \left(m_t^2+m_t\sqrt{m_W^2+\beta_t^2(m_t^2-m_W^2)}\right)\,,
    \nn \\ 
    u&=m_t^2-\frac{1-\beta_t\cos\theta}{1-\beta_t^2} \left(m_t^2+m_t\sqrt{m_W^2+\beta_t^2(m_t^2-m_W^2)}\right)\,.
\end{align}
In our calculation, for each phase space point $(\beta_t,\cos\theta)$,
we first compute the above $s,t$ and $u$ variables in high precision floating number, e.g., by keeping over 20 digits, 
and then transform them to rational numbers
before inserting them to the expressions of the coefficients of the master integrals.
We have observed a strong cancellation among different terms in each coefficient  for small or large $\beta_t$.
The cancellation of all the divergences in eq.(\ref{eq:fin}), either analytically or numerically,  is an important check of the whole calculation.

\begin{figure}
    \centering
    \includegraphics[width = 0.49 \textwidth]{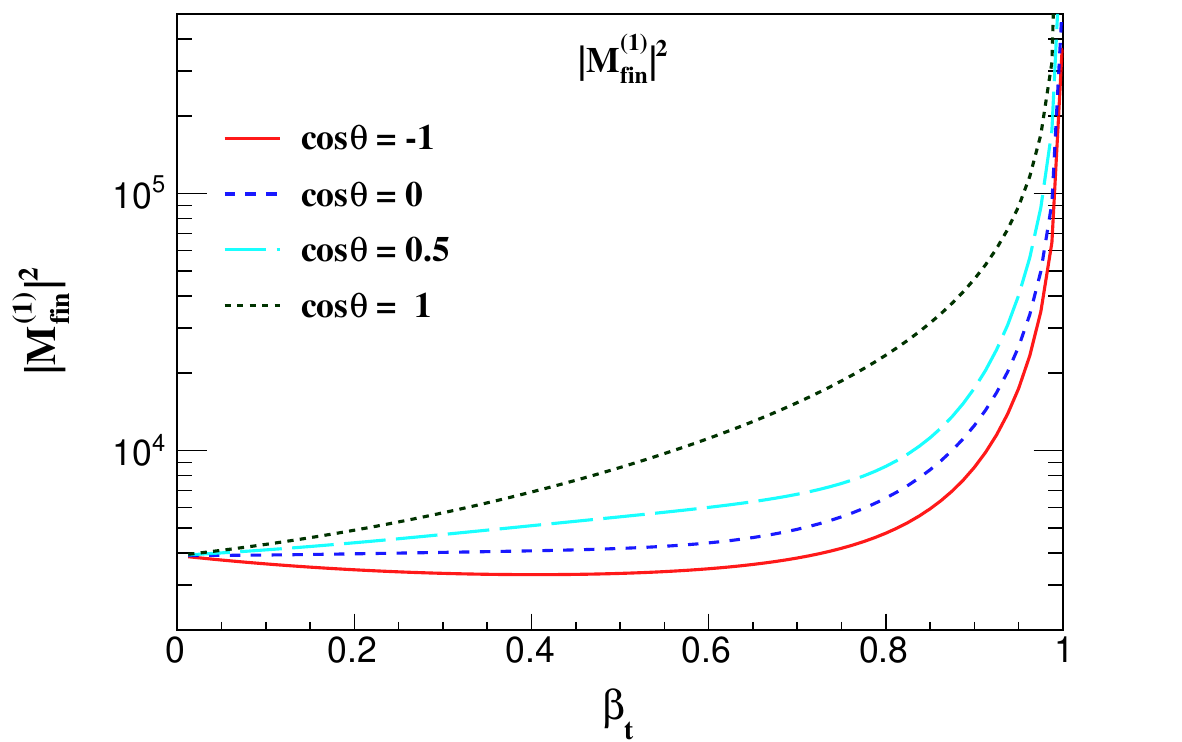}
    \includegraphics[width = 0.49 \textwidth]{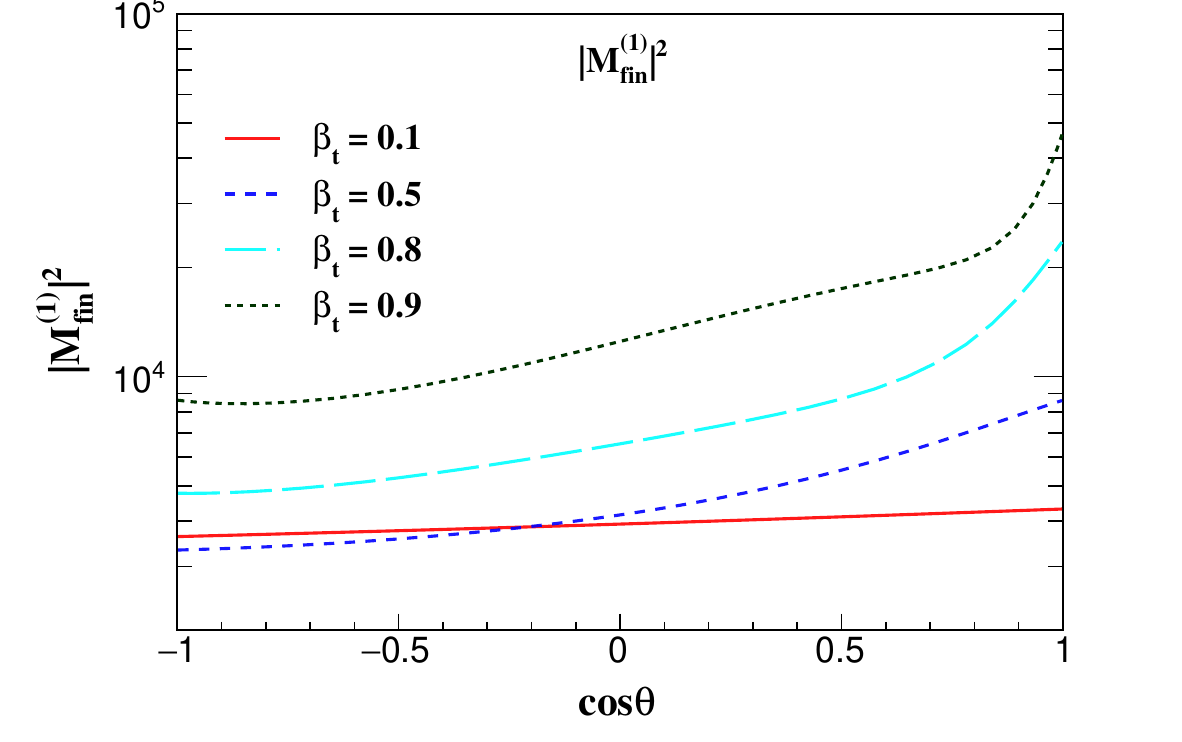}
    \caption{$|\mathcal{M}_{\rm fin}^{(1)}|^2$ with fixed $\cos\theta$ (left) and $\beta_t$ (right). The other parameters are set the same as figure~\ref{fig:bare}.}
    \label{fig:finite}
\end{figure}

\begin{figure}
    \centering
    \includegraphics[width = 0.49 \textwidth]{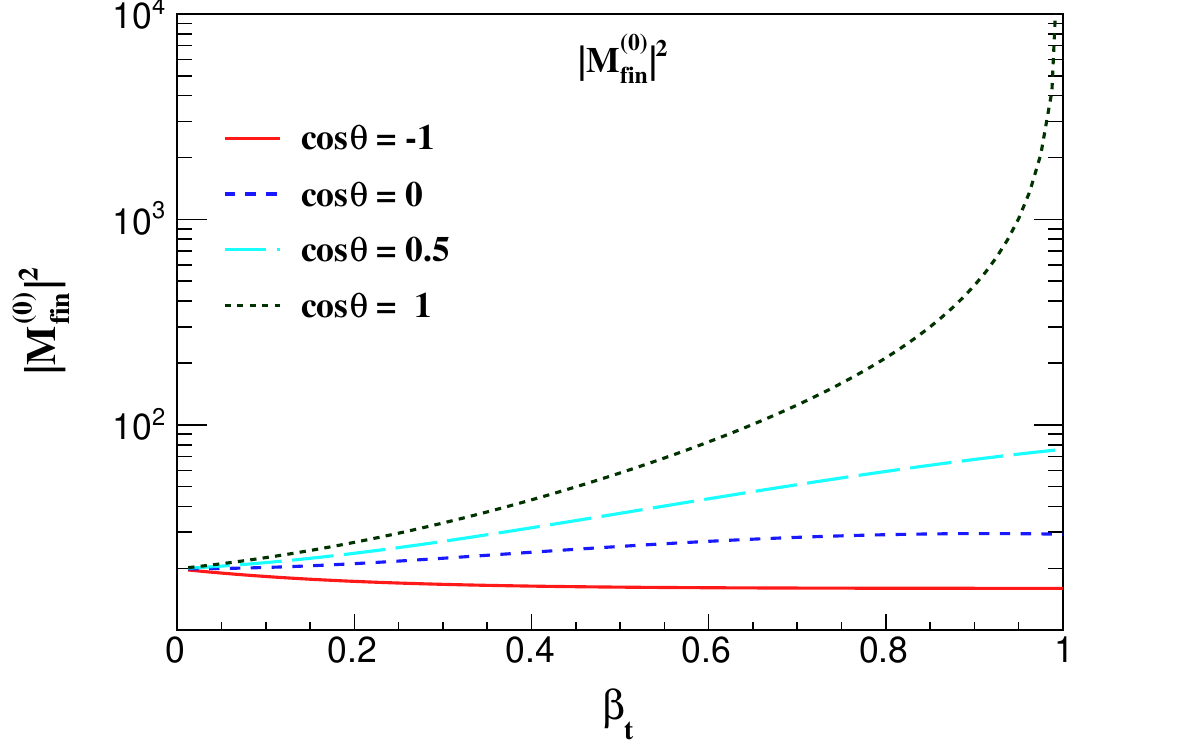}    
    \includegraphics[width = 0.49 \textwidth]{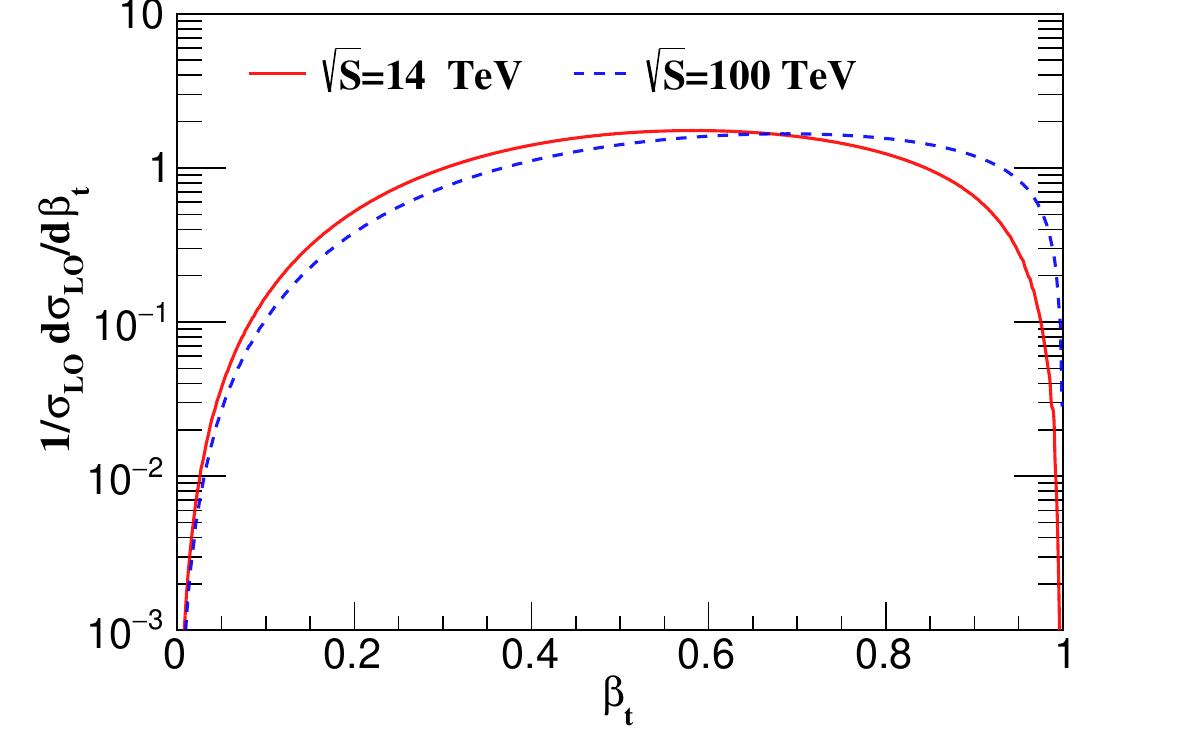}       
    \caption{The LO squared amplitude (left) and normalized cross section (right) as a function of $\beta_t$. In the left plot, the other parameters  are set the same as figure~\ref{fig:bare}. In the right plot, the renormalization and factorization scales are set to be $m_t$. The red and blue lines correspond to the collider energy $\sqrt{S}=14$ and 100 TeV, respectively, while $\beta_t$ is calculated in partonic central-of-mass frame.  }
    \label{fig:lo}
\end{figure}

At last, we turn to the finite part.
Figure~\ref{fig:finite} shows the numerical result of $\Big| \mathcal{M}^{(1)}_{\rm fin} \Big|^2$ as a function of $\beta_t$ and $\cos\theta$.
It can be seen that the finite one-loop squared amplitude increases remarkably when $\beta_t$ becomes large.
The reason is that  the mass of the top quark becomes vanishing in the limit $\beta_t\to 1$ \footnote{Or one may think that it is the region where the collider energy goes to infinity. This does not affect the discussion on the divergences. }.
New collinear divergences, which have been regularized by $m_t$ and appeared as logarithms of  $ 1-\beta_t^2$, will appear.
From the right plot of figure~\ref{fig:finite}, we observe that the finite part is insensitive to $\cos\theta$ when $\beta_t$ is small.
But the dependence of the finite part on 
$\cos\theta$ becomes stronger with the increasing of $\beta_t$.

For comparison, we present the LO squared amplitude in the left plot in figure \ref{fig:lo}.
We see that it is divergent  only in the limit $\beta_t\to 1$ and $\cos\theta\to 1$.
This is the region when the $t$-channel propagator in figure  \ref{fig:LO} becomes on-shell.
Notice that the two kinds of divergences in the limit $\beta_t\to 1$ are different, because the collinear divergences discussed above can appear for any values of $\cos\theta$.
In practice, both of them are not harmful to the hadronic cross section  due to the suppression of parton distribution function in the large $\beta_t$ region.
After phase space integration,  the LO partonic cross section is proportional to $\ln (1-\beta_t)$, 
but the hadronic cross section is vanishing as $\beta_t\to 1$,
as shown in the right plot of  figure \ref{fig:lo}.
This means that the suppression of the parton distribution function surpasses the enhancement of the partonic cross section.
This is also the case at higher orders, since only $\ln^n(1-\beta_t)$  can appear in the partonic cross section.
From the right plot of figure \ref{fig:lo}, we also observe that the suppression becomes weaker with the increasing of the collider energy.

\begin{figure}
    \centering
    \includegraphics[width = 0.49 \textwidth]{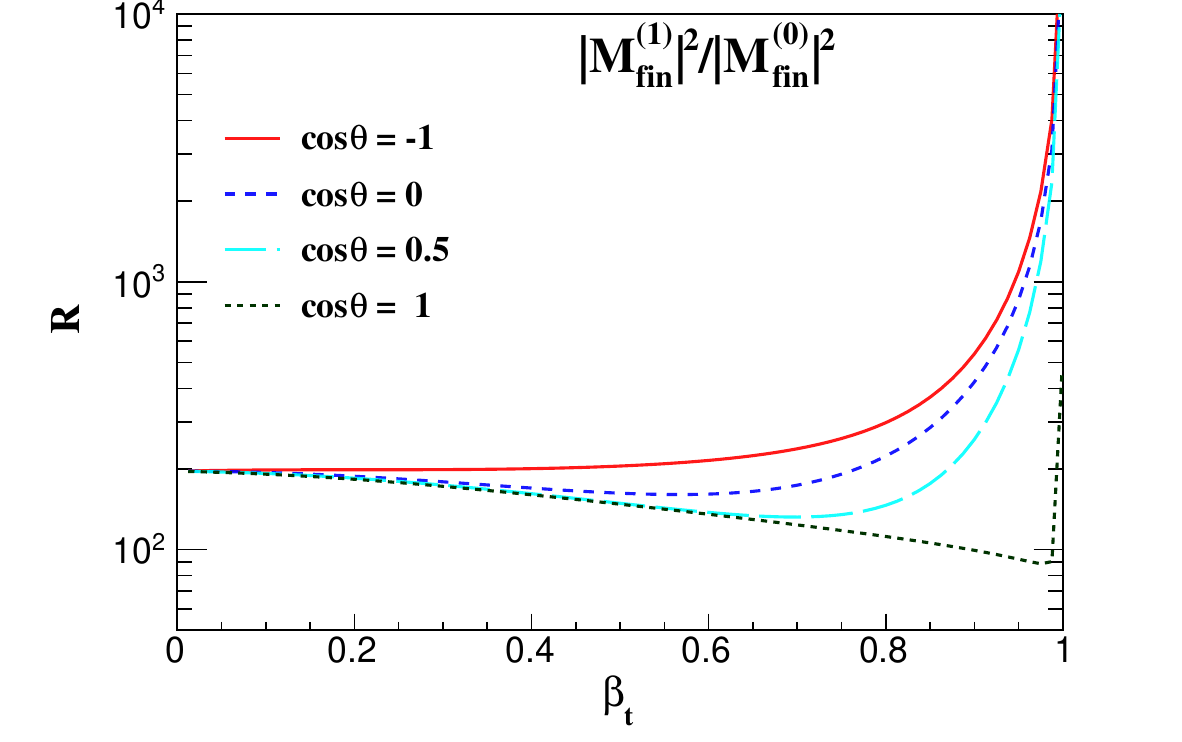} 
    \includegraphics[width = 0.49 \textwidth]{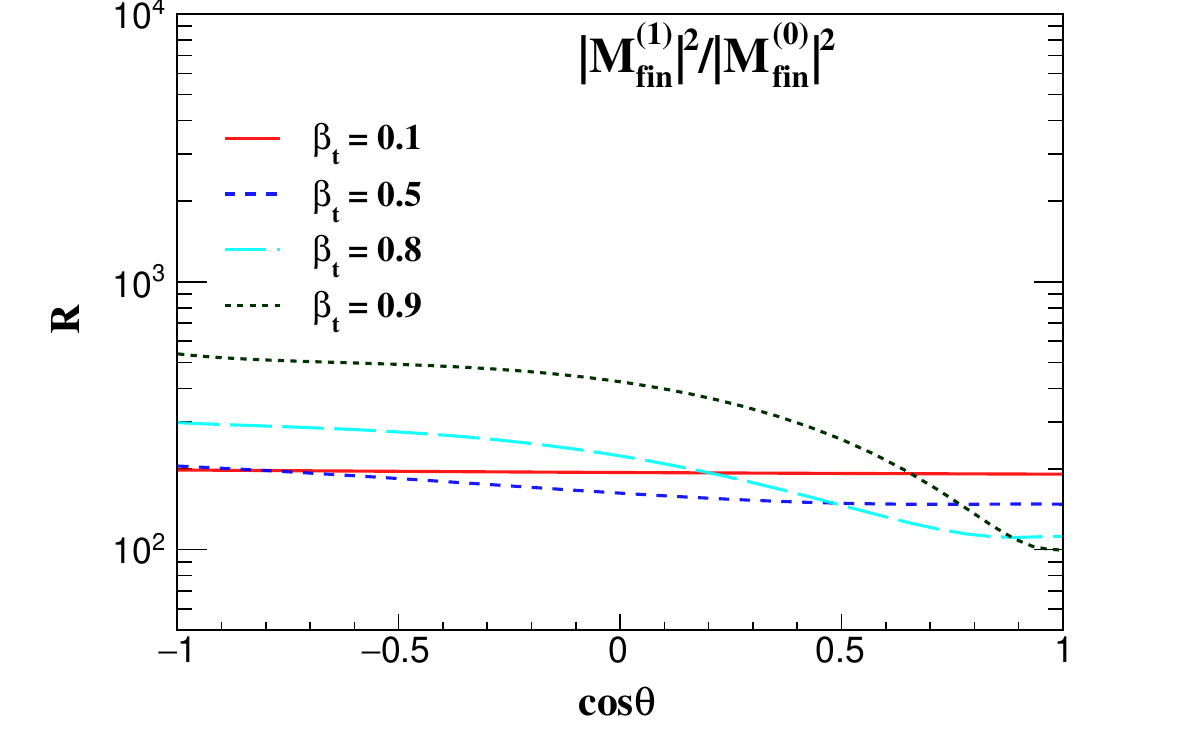}    
    \caption{The ratio of $|\mM_{\rm fin}^{(1)}|^2$ to $|\mM_{\rm fin}^{(0)}|^2$ as a function of $\beta_t$ (left plot) and $\cos\theta$ (right plot). }
    \label{fig:ratio}
\end{figure}

To see the impact of the one-loop amplitude  squared, we show the ratio $R$ of $|\mM_{\rm fin}^{(1)}|^2$ to $|\mM_{\rm fin}^{(0)}|^2$ in figure \ref{fig:ratio}.
We find that the ratio $R$ changes slowly away from the region of $\beta_t\to 1$ or $\cos\theta\to 1$ and grows dramatically for large $\beta_t$.
For the phase space with $\beta_t<0.9$, which is the dominant region contributing to a real collider process, the corrections from one-loop amplitude squared are at the percent level, if the strong coupling is taken to be around $0.1$.

\section{Expansion in $m_W^2$}
\label{sec:mWexp}

The complexity of the loop amplitude mounts up prominently 
as the number of the independent scales increases.
There are four independent scales in the process of $tW$ production, i.e., $m_t^2, m_W^2, s, t$. 
It is beneficial to computation if one can expand the amplitude in one scale.
For the process we are interested in,
setting $m_W=0$ does not bring any new IR divergences over the full phase space 
due to the massive top quark propagator.
Therefore, it is feasible to take the Taylor series expansion of the amplitude with respect to $m_W^2$.
After setting $m_W=0$, one only needs to deal with integrals depending on three variables, i.e., $s,t,m_t^2$.
As a consequence,  both the IBP reduction and calculation of master integrals are simplified considerably. The square roots in eq.(\ref{eq:MIs}) will disappear, and thus the one-loop results become easier. 
Furthermore, at two-loop level, small mass expansion can make the calculation of some master integrals that would contain elliptic functions when keeping finite $m_W$ possible in terms of multiple polylogarithms. 
A similar method has been adopted in the analytic  calculation of the two-loop $gg\to HH$ amplitude \cite{Xu:2018eos,Wang:2020nnr}.

Given that the analytic result is still out of reach,
we want to apply the method of Taylor expansion in $m_W^2$ to compute the approximate two-loop amplitude of $tW$ production.
In this section, we demonstrate the validity of this method using the exact one-loop result we have obtained.  As shown in eq.(\ref{eq:mlo}), the LO scattering amplitude contains the denominator $s$ and $u-m_t^2$. One reasonable choice is to keep the invariant $s$ and $u$ and replace $t\to -s-u+m_t^2+m_W^2$ before expanding the amplitudes in a series of $m_W^2$.

Regarding the calculation of master integrals, we explicitly show the expansion of $I^i_{n_1,n_2,n_3,n_4}$ defined in eq.(\ref{def}), 
\begin{eqnarray}
	I^i_{n_1,n_2,n_3,n_4}\left(s,u,m_W^2,m_t^2\right) =  \sum_{n=0}^{\infty}\frac{(m_W^2)^n}{n!}\left.\frac{\partial^n I^i_{n_1,n_2,n_3,n_4}}{\partial (m_W^2)^n}\right|_{m_W^2=0}.
	\label{eq:Iexpansion}
\end{eqnarray}
The differential operator of $m_W^2$ should be written in a form that can directly act on a loop integral,
\begin{eqnarray}
\frac{\partial}{\partial(m_W^2)} = \left(c_1 k_1 + c_2 k_2 + c_3 k_3\right)\cdot \frac{\partial}{\partial k_3}
\label{eq:partialk}
\end{eqnarray}
with the coefficients
\begin{eqnarray}
c_1 &=& \frac{u \left(-m_W^2+2 s+u\right)+m_t^2 \left(m_W^2-u\right)}{2 s \left(u \left(-m_W^2+s+u\right)+m_t^2 \left(m_W^2-u\right)\right)},\nonumber\\
c_2 &=& \frac{\left(u-m_t^2\right) \left(-m_t^2+s+u\right)}{2 s \left(u \left(-m_W^2+s+u\right)+m_t^2 \left(m_W^2-u\right)\right)},\nonumber\\
c_3 &=& \frac{m_t^2-u}{2 \left(u \left(-m_W^2+s+u\right)+m_t^2 \left(m_W^2-u\right)\right)}.
\end{eqnarray}
Notice that these $c_i$ coefficients should be expanded further in a series of $m_W^2$ before taking the limit $m_W^2\to 0$ in eq.(\ref{eq:Iexpansion}).

Actually, this differential operators can act not only on the master integrals, but also on the whole squared amplitude.
However, we do not apply them to the variable $m_W^2$ in the summation of the polarizations of the $W$ boson in eq.(\ref{eq:pol-sum}).
This denominator arises because the longitudinal polarization state of the $W$ boson has a different coupling to quarks from the transverse ones.
If we separate the contribution from different polarizations and express the coupling of
the longitudinal  polarization state and quarks in terms of the top quark Yukawa coupling, there is no such $m_W^2$ factor in the denominator, according to the Goldstone boson equivalence theorem \cite{Cornwall:1974km,Lee:1977eg}.
This property can also be seen if we use Feynman-'t Hooft gauge to sum the polarizations of the $W$ boson 
\begin{align}
    \sum_i \epsilon_i^{*\mu}(k_3)\epsilon_i^{\nu}(k_3) = -g^{\mu\nu}
\end{align}
and add the contribution of the Goldstone boson production diagram, which does not involve a polarization sum.

After applying the differential operator to master integrals, each term in the series of $m_W^2$ can be expressed as a linear combination of scalar integrals. Then we reduce them to a set of master integrals with $m_W=0$.
The advantages of the expansion come from two aspects.
First, the number of these master integrals is less than that before expansion. 
Second, the analytic computation is easier to perform.
For example, the square root in eq.(\ref{eq:MIs}) does not exist at $m_W=0$.
Further, the set of master integrals is the same at all orders in the expansion,
therefore no additional computation cost needs to be paid if we want to obtain the result of even higher orders in the expansion of $m_W^2$ in principle.

\begin{figure}
    \centering
    \includegraphics[width = 0.99 \textwidth]{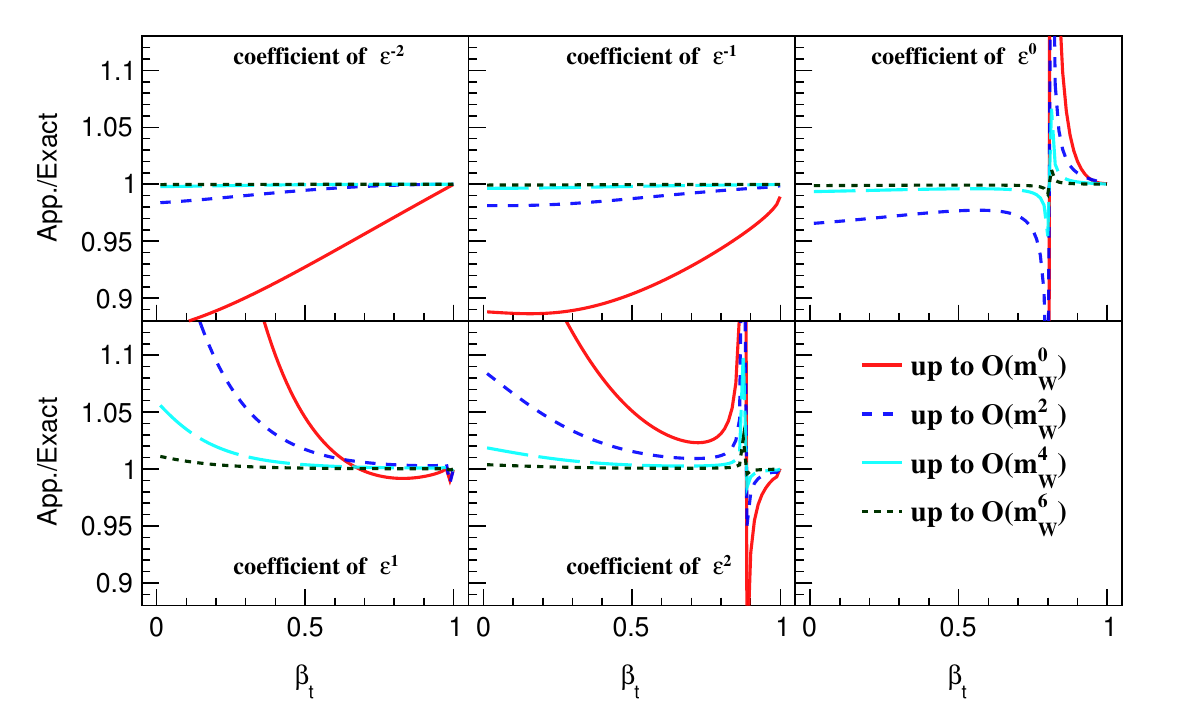}
    \vspace{-0.5cm}
    \caption{The accuracy of $m_W$ expansion for the real part of  one-loop integral $I^1_{1,1,1,1}$ up to $\epsilon^2$ and $\mathcal{O}(m_W^6)$ with $\cos\theta=0.5$. The red, blue, cyan and dark green lines stand for the expansions up to $\mathcal{O}(m_W^0)$, $\mathcal{O}(m_W^2)$, $\mathcal{O}(m_W^4)$, and $\mathcal{O}(m_W^6)$, respectively.
    The notable divergences in the plots appear because the exact results almost vanish around that specific value of $\beta_t$.}
    \label{fig:D01}
\end{figure}

Figure~\ref{fig:D01} demonstrates the accuracy of the $m_W$ expansion, taking the real part of $I^1_{1,1,1,1}$ as an example.
We compare the expanded results for the coefficients of $\ep^i$ with $i=-2,\cdots, 2$ to the exact ones.
As more higher orders in $m_W^2$ are included,
the  expansion approximation becomes better.
The difference between the expansion up to $\mathcal{O}(m_W^6)$ and the exact values are  at the permille level or better. 
We also find that expansion series converge faster in the larger $\beta_t$ region.
This is reasonable because the expansion parameter $m_W^2/s$ is generally smaller in this region.
In the plots, there are some regions where the curves seem to be divergent. 
We have checked the absolute values of both the expanded and exact results, and found that they are almost vanishing in these regions.

\begin{figure}
    \centering
    \includegraphics[width = 0.49 \textwidth]{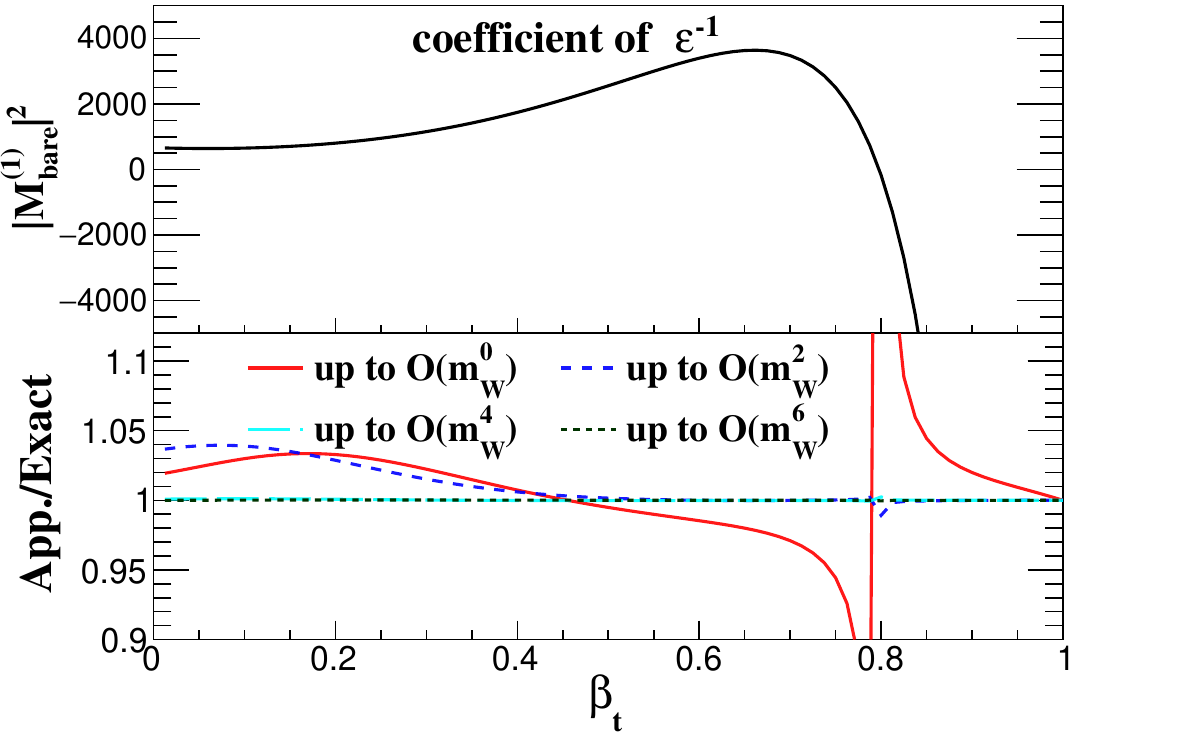}
    \includegraphics[width = 0.49 \textwidth]{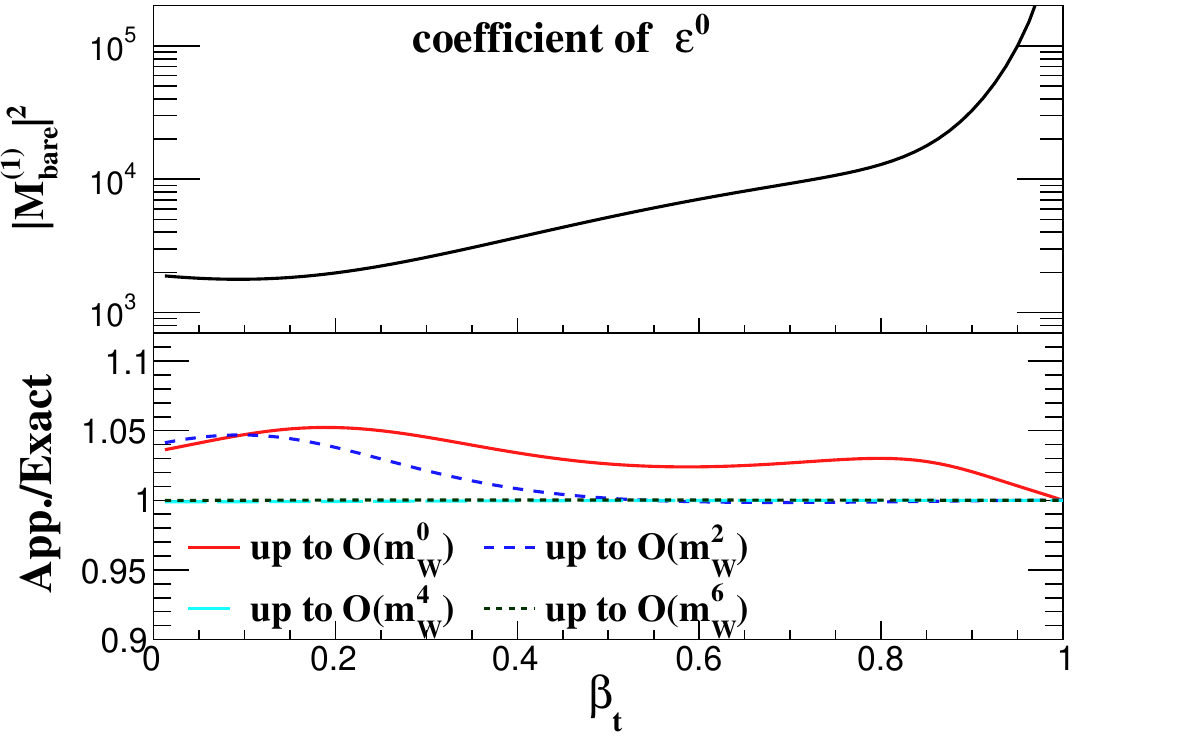}
    \caption{The numerical results for the coefficients of $\epsilon^{-1}$ (left) and $\epsilon^0$ (right) in $|\mathcal{M}_{\rm bare}^{(1)}|^2$ and their expansions in $m_W$ with $\cos\theta=0.5$ as a function of $\beta_t$. 
    The upper panels shows the exact results of one-loop squared amplitudes. The lower panels show the ratios between the $m_W$ expansions  and the exact results.  The color scheme is the same as in figure~\ref{fig:D01}.}
    \label{fig:mwExp}
\end{figure}

Then we consider the expansion of the squared amplitudes.
As mentioned above, the differential operator can be applied to the squared amplitude. This can be done just after creating the amplitude from the Feynman diagrams.
Firstly, we replace $k_4$ by $k_1+k_2-k_3$ everywhere.
Secondly,  we contract the Lorentz indices in the spin summed squared amplitude.
Thirdly, we apply the differential operator in eq.(\ref{eq:partialk}) to the squared amplitude\footnote{The differential operators must also be applied to the polarization sum of the $W$ boson except for the denominator $m_W^2$.},
and reduce all the scalar integrals to master integrals with $k_3^2=0$.
However, we will use another method. Since we have expressed the squared amplitude as 
a combination of the master integrals and their coefficients,
we only have to expand the coefficients in $m_W^2$, given that the master integrals have been expanded as discussed above.
Figure~\ref{fig:mwExp} shows $|\mathcal{M}_{\rm bare}^{(1)}|^2$ and their expansions in $m_W^2$ with $\cos\theta=0.5$ as a function of $\beta_t$. It can be seen that the differences between the exact results and the expanded ones up to $\mathcal{O}(m_W^4)$ are already notably small. 
The corrections of order $m_W^6$ are negligible.

We have also tried to expand only all the master integrals but keep the exact form of the coefficients of the master integrals.
One may expect that this approach would have a better approximation to the exact result.
On the contrary, we find that the expansion is worse, sometimes even seems divergent in small $\beta_t$ regions.
As mentioned before, the coefficients contain a lot of terms, in which the powers of the kinematic variables are very high. 
For specific values of the kinematic variables, the cancellation among these terms happens over about ten digits.
In the expansion in $m_W^2$, this cancellation holds at each order.  
If the coefficients contain higher orders in $m_W^2$ than the master integrals, the cancellation is not complete. 
As a result, the deviation between the expanded results and the exact ones  may be large.

\section{Conclusions}
\label{sec:conclusion}

The precise prediction of $tW$ associated production at a hadron collider is important to study the properties of the top quark and the $W$ boson.
We have computed the analytic results for the one-loop squared amplitudes for this process using canonical differential equations.
The renormalized amplitude squared has up to $\ep^{-4}$ poles,
which have been checked against the general infrared structures predicted by anomalous dimensions.
The finite part gives rise to about  a few percent corrections compared to the corresponding LO results. 
We also investigate how to obtain approximated results using the method of expansion in $m_W^2$,
and find that the expanded results up to $\mO(m_W^4)$ are already accurate enough to present the exact results.
It is promising to apply this expansion method in calculating the interference of two-loop and tree level amplitudes.

\section*{Acknowledgements}
This work was supported in part by the National Science Foundation of China under grant No.12175048, No.12005117, No.12147154 and No.12075251.  The work of L.D. and J.W. was also supported by the Taishan Scholar Foundation of Shandong province (tsqn201909011).

\appendix
\section{Rational matrices and boundary conditions for the second integral family}
\label{sec:diffB}

The canonical basis of the second family consists of
\bqa
N_1&=&\epsilon\, I^2_{0,0,0,2}\,,\nonumber\\
N_2&=&\epsilon\, t I^2_{0,1,0,2}\,,\nonumber\\
N_3&=&\epsilon\, u I^2_{1,0,2,0}\,,\nonumber\\
N_4&=&\epsilon\, m_W^2 I^2_{0,1,2,0}\,,\nonumber\\
N_5&=&\epsilon^2\, (u-m_t^2) I^2_{1,0,1,1}\,,\nonumber\\
N_6&=&\epsilon^2\, (t-m_W^2)I^2_{0,1,1,1}\,,\nonumber\\
N_7&=&\epsilon^2\, (t-m_t^2)(u-m_t^2)I^2_{1,1,1,1}\,,
\eqa
which satisfy the differential equation
\bqa
d\, \text{{\bf N}}(y,z,v;\epsilon)=\epsilon\, (d \, \tilde{B})\,  \text{{\bf N}}(y,z,v;\epsilon)
\eqa
with
\bqa
d\, \tilde{B}=\sum_{i=1}^{10} P_i\,  d \ln(L_i).
\eqa
  The letters are given by 
\begin{align}
\begin{alignedat}{2}
L_1 & =y\,,&\quad
L_2 & =v\,, \\
L_3 & =1+z^2-y-v\,, &\quad
L_4 & =z^2\,, \\
L_5 & =y-1\,,&\quad
L_6 & =v-1\,,\\
L_7 & =z^2-1 \,,&\quad
L_8 & =y-z^2 \, , \\
L_9 & =v-z^2\,,&\quad
L_{10} & =y\, v -z^2
\end{alignedat}
\stepcounter{equation}\tag{\theequation}
\label{alphabet}
\end{align}
with $v\equiv\frac{u}{m_t^2}$ and $y,z$ defined in eq.(\ref{xyzd}).
The rational matrices $P_i$ are 
\bqa
P_1 &=& \left(
\begin{array}{ccccccc}
 0 & 0 & 0 & 0 & 0 & 0 & 0 \\
 0 & 1 & 0 & 0 & 0 & 0 & 0 \\
 0 & 0 & 0 & 0 & 0 & 0 & 0 \\
 0 & 0 & 0 & 0 & 0 & 0 & 0 \\
 0 & 0 & 0 & 0 & 0 & 0 & 0 \\
 0 & 1 & 0 & 0 & 0 & 0 & 0 \\
 0 & 0 & 0 & 0 & 0 & 0 & 0 \\
\end{array}
\right),~~
P_2=\left(
\begin{array}{ccccccc}
 0 & 0 & 0 & 0 & 0 & 0 & 0 \\
 0 & 0 & 0 & 0 & 0 & 0 & 0 \\
 0 & 0 & 1 & 0 & 0 & 0 & 0 \\
 0 & 0 & 0 & 0 & 0 & 0 & 0 \\
 0 & 0 & 1 & 0 & 0 & 0 & 0 \\
 0 & 0 & 0 & 0 & 0 & 0 & 0 \\
 0 & 0 & 0 & 0 & 0 & 0 & 0 \\
\end{array}
\right),\nonumber\\
P_3 &=& \left(
\begin{array}{ccccccc}
 0 & 0 & 0 & 0 & 0 & 0 & 0 \\
 0 & 0 & 0 & 0 & 0 & 0 & 0 \\
 0 & 0 & 0 & 0 & 0 & 0 & 0 \\
 0 & 0 & 0 & 0 & 0 & 0 & 0 \\
 0 & 0 & 0 & 0 & 0 & 0 & 0 \\
 0 & 0 & 0 & 0 & 0 & 0 & 0 \\
 -\frac{1}{2} & -1 & -1 & 1 & 1 & 1 & 1 \\
\end{array}
\right),~~
P_4= \left(
\begin{array}{ccccccc}
 0 & 0 & 0 & 0 & 0 & 0 & 0 \\
 0 & 0 & 0 & 0 & 0 & 0 & 0 \\
 0 & 0 & 0 & 0 & 0 & 0 & 0 \\
 0 & 0 & 0 & 1 & 0 & 0 & 0 \\
 0 & 0 & 0 & 0 & 0 & 0 & 0 \\
 0 & 0 & 0 & -1 & 0 & 0 & 0 \\
 0 & 0 & 0 & 0 & 0 & 0 & 0 \\
\end{array}
\right),\nonumber\\
P_5 &=& \left(
\begin{array}{ccccccc}
 0 & 0 & 0 & 0 & 0 & 0 & 0 \\
 -1 & -2 & 0 & 0 & 0 & 0 & 0 \\
 0 & 0 & 0 & 0 & 0 & 0 & 0 \\
 0 & 0 & 0 & 0 & 0 & 0 & 0 \\
 0 & 0 & 0 & 0 & 0 & 0 & 0 \\
 0 & 0 & 0 & 0 & 0 & 0 & 0 \\
 0 & 0 & 0 & 0 & 0 & 0 & -2 \\
\end{array}
\right),~~
P_6= \left(
\begin{array}{ccccccc}
 0 & 0 & 0 & 0 & 0 & 0 & 0 \\
 0 & 0 & 0 & 0 & 0 & 0 & 0 \\
 -1 & 0 & -2 & 0 & 0 & 0 & 0 \\
 0 & 0 & 0 & 0 & 0 & 0 & 0 \\
 0 & 0 & 0 & 0 & 0 & 0 & 0 \\
 0 & 0 & 0 & 0 & 0 & 0 & 0 \\
 0 & 0 & 0 & 0 & 0 & 0 & -2 \\
\end{array}
\right),\nonumber\\
P_7 &=& \left(
\begin{array}{ccccccc}
 0 & 0 & 0 & 0 & 0 & 0 & 0 \\
 0 & 0 & 0 & 0 & 0 & 0 & 0 \\
 0 & 0 & 0 & 0 & 0 & 0 & 0 \\
 -1 & 0 & 0 & -2 & 0 & 0 & 0 \\
 0 & 0 & 0 & 0 & 0 & 0 & 0 \\
 0 & 0 & 0 & 0 & 0 & 0 & 0 \\
 1 & 0 & 0 & 2 & 0 & 0 & 0 \\
\end{array}
\right),~~
P_8=\left(
\begin{array}{ccccccc}
 0 & 0 & 0 & 0 & 0 & 0 & 0 \\
 0 & 0 & 0 & 0 & 0 & 0 & 0 \\
 0 & 0 & 0 & 0 & 0 & 0 & 0 \\
 0 & 0 & 0 & 0 & 0 & 0 & 0 \\
 0 & 0 & 0 & 0 & 0 & 0 & 0 \\
 0 & 0 & 0 & 0 & 0 & 0 & 0 \\
 0 & 2 & 0 & -2 & 0 & 0 & 0 \\
\end{array}
\right),\nonumber\\
P_9 &=& \left(
\begin{array}{ccccccc}
 0 & 0 & 0 & 0 & 0 & 0 & 0 \\
 0 & 0 & 0 & 0 & 0 & 0 & 0 \\
 0 & 0 & 0 & 0 & 0 & 0 & 0 \\
 0 & 0 & 0 & 0 & 0 & 0 & 0 \\
 0 & 0 & 0 & 0 & 0 & 0 & 0 \\
 0 & 0 & 0 & 0 & 0 & 0 & 0 \\
 0 & 0 & 2 & -2 & 0 & 0 & 0 \\
\end{array}
\right),~~
P_{10}=\left(
\begin{array}{ccccccc}
 0 & 0 & 0 & 0 & 0 & 0 & 0 \\
 0 & 0 & 0 & 0 & 0 & 0 & 0 \\
 0 & 0 & 0 & 0 & 0 & 0 & 0 \\
 0 & 0 & 0 & 0 & 0 & 0 & 0 \\
 0 & 0 & 0 & 0 & 0 & 0 & 0 \\
 0 & 0 & 0 & 0 & 0 & 0 & 0 \\
 -\frac{1}{2} & -1 & -1 & 1 & -1 & -1 & 1 \\
\end{array}
\right).
\eqa

The boundaries of these basis integrals 
are fixed by either direct computation (for $N_1$)
or regular conditions (for $N_2\sim N_7$).
More explicitly,
$N_2,\cdots,N_7$ are regular at $t=0,u=0,m_W=0,u=m_t^2,t=m_W^2,s=0$, without any singularity,
respectively.

\section{Perturbative expansion of renormalization factors}
\label{sec:renormfactor}

The strong coupling $\alpha_s$ is renomalized in the $\overline{\rm MS}$ scheme.
The renormalization factor is given by
\cite{Caswell:1974gg,Jones:1974mm}
\begin{align}
     Z_{\alpha_s}=1-\frac{\alpha_s}{4\pi}
   \frac{\beta_0}{ \ep}
    +\left(\frac{\alpha_s}{4\pi} \right)^2 \left(\frac{\beta_0^2}{\ep^2}-\frac{\beta_1}{2\ep}\right)+\mO(\alpha_s^3)
\end{align}
with 
\begin{align}
   \beta_0 & = \frac{11}{3}C_A-\frac{4}{3}T_F n_f\,, \nn \\
   \beta_1 & = \frac{34}{3}C_A^2 - \frac{20}{3} C_A T_F n_f - 4 C_F T_F n_f \,.
\end{align}
In QCD, we have $C_A=3,C_F=4/3,T_F=1/2$.
Here $n_f$ is the total number of quark flavors, including 
both massless and massive quarks.
As a result, the running strong coupling satisfies the renomalization equation
\begin{align}
\frac{d\alpha_s}{d\ln \mu} = 
-2\alpha_s\left(\frac{\alpha_s}{4\pi}\beta_0+\left(\frac{\alpha_s}{4\pi}\right)^2\beta_1+\cdots\right)\,.
\end{align}

In our calculation, the wave function renormalization factors are determined by the on-shell renormalization condition for the external particles,
and have been expanded in a series of $\alpha_s$.
The first three orders are given by
\cite{Broadhurst:1991fy,Czakon:2007wk,Czakon:2007ej}
\begin{align}
    Z_2^g=&1+\left(\frac{\alpha_s}{4\pi}\right)T_F n_h D_\epsilon\left(-\frac{4}{3\epsilon}\right) +\left(\frac{\alpha_s}{4\pi}\right)^2 T_F n_h  D_\epsilon^2\Biggl[ C_F\left(-\frac{2}{\epsilon}-15\right)+C_A\left(\frac{35}{9\epsilon^2}-\frac{5}{2\epsilon}\right.\nn \\
    &\left.+\frac{13}{12}-\frac{11\pi^2}{27}\right)+T_F\left(-\frac{16}{9\epsilon^2}n_l+\frac{4\pi^2}{27}n_f\right)+\frac{4}{3}\beta_0\ln\left(\frac{\mu^2}{m_t^2}\right) \left(-\frac{1}{\epsilon}+\frac{1}{2}\ln\left(\frac{\mu^2}{m_t^2}\right)\right) \Biggr],\nn \\
    Z^b_2 =& 1+\left(\frac{\alpha_s}{4\pi}\right)^2 C_F T_F n_h \left(\frac{1}{\ep} + 2\ln \frac{\mu^2}{m_t^2} - \frac{5}{6}\right) , \nn \\    
    Z_2^t=&1+\left(\frac{\alpha_s}{4\pi}\right) C_F D_\epsilon \left(-\frac{3}{\epsilon}-4-8\epsilon-16\epsilon^2\right) + \left(\frac{\alpha_s}{4\pi}\right)^{2} C_F D_\epsilon^2\Bigg[T_F n_h \left(\frac{1}{\epsilon}+\frac{947}{18}-5\pi^2\right) \nn \\
    &+ T_F n_l \left(-\frac{2}{\epsilon^2}+\frac{11}{3\epsilon}+\frac{113}{6}+\frac{5\pi^2}{3}\right)+ C_F \bigg(\frac{9}{2\epsilon^2}+\frac{51}{4\epsilon}+\frac{433}{8}-13\pi^2 \nn \\
    &+16\pi^2\ln2-24\zeta(3)\bigg)+C_A\left(\frac{11}{2\epsilon^2}-\frac{127}{12\epsilon}-\frac{1705}{24}+\frac{49\pi^2}{12}-8\pi^2\ln2+12\zeta(3)\right) \nn \\
    &+ \beta_0\ln\left(\frac{\mu^2}{m_t^2}\right)\left(-\frac{3}{\epsilon}-4+\frac{3}{2}\ln\left(\frac{\mu^2}{m_t^2}\right)\right)\Bigg],\nn \\
    Z_m^t=&1+\left(\frac{\alpha_s}{4\pi}\right) C_F D_\epsilon \left(-\frac{3}{\epsilon}-4-8\epsilon-16\epsilon^2\right) + \left(\frac{\alpha_s}{4\pi}\right)^{2} C_F D_\epsilon^2\Bigg[T_F n_h \left(-\frac{2}{\epsilon^2}+\frac{5}{3\epsilon}+\frac{143}{6}\right.\nn \\
    &\left.-\frac{7\pi^2}{3}\right)+T_F n_l \left(-\frac{2}{\epsilon^2}+\frac{5}{3\epsilon}+\frac{71}{6}+\frac{5\pi^2}{3}\right)+C_F \biggl(\frac{9}{2\epsilon^2}+\frac{45}{4\epsilon}+\frac{199}{8}-5\pi^2 \nn \\
    &+8\pi^2\ln2-12\zeta(3)\biggr)+C_A\left(\frac{11}{2\epsilon^2}-\frac{97}{12\epsilon}-\frac{1111}{24}+\frac{5\pi^2}{12}-4\pi^2\ln2+6\zeta(3)\right) \nn \\
    &+\beta_0\ln\left(\frac{\mu^2}{m_t^2}\right)\left(-\frac{3}{\epsilon}-4+\frac{3}{2}\ln\left(\frac{\mu^2}{m_t^2}\right)\right)\Bigg]
\end{align}
with 
\begin{align}
     D_{\ep} \equiv \frac{ \Gamma(1+\ep)}{e^{-\gamma_E\ep}} \left(\frac{ \mu^2}{m_t^2}\right)^{\ep}.
\end{align}
We have used $n_h~(n_l)$ to denote the number of massive (massless) quarks, though
in our case it is natural to set $n_h=1$.
Notice that $Z^t_2$ and $Z^t_m$ are the same at $\mO(\alpha_s)$ if we do not distinguish the IR and UV divergences in dimensional regularization,
but not the same at higher orders \cite{Melnikov:2000zc}.

\bibliography{tw-oneloop-squared}

\providecommand{\href}[2]{#2}\begingroup\raggedright\begin{thebibliography}{10}

\bibitem{CDF:2022hxs}
{\scshape CDF} collaboration, T.~Aaltonen et~al., \emph{{High-precision
  measurement of the W boson mass with the CDF II detector}},
  \href{https://doi.org/10.1126/science.abk1781}{\emph{Science} {\bfseries 376}
  (2022) 170--176}.

\bibitem{Aad:2012xca}
{\scshape ATLAS} collaboration, G.~Aad et~al., \emph{{Evidence for the
  associated production of a $W$ boson and a top quark in ATLAS at $\sqrt{s}=7$
  TeV}}, \href{https://doi.org/10.1016/j.physletb.2012.08.011}{\emph{Phys.
  Lett.} {\bfseries B716} (2012) 142--159},
  [\href{https://arxiv.org/abs/1205.5764}{{\ttfamily 1205.5764}}].

\bibitem{Aad:2015eto}
{\scshape ATLAS} collaboration, G.~Aad et~al., \emph{{Measurement of the
  production cross-section of a single top quark in association with a $W$
  boson at 8 TeV with the ATLAS experiment}},
  \href{https://doi.org/10.1007/JHEP01(2016)064}{\emph{JHEP} {\bfseries 01}
  (2016) 064}, [\href{https://arxiv.org/abs/1510.03752}{{\ttfamily
  1510.03752}}].

\bibitem{Aaboud:2016lpj}
{\scshape ATLAS} collaboration, M.~Aaboud et~al., \emph{{Measurement of the
  cross-section for producing a W boson in association with a single top quark
  in pp collisions at $ \sqrt{s}=13 $ TeV with ATLAS}},
  \href{https://doi.org/10.1007/JHEP01(2018)063}{\emph{JHEP} {\bfseries 01}
  (2018) 063}, [\href{https://arxiv.org/abs/1612.07231}{{\ttfamily
  1612.07231}}].

\bibitem{Aaboud:2017qyi}
{\scshape ATLAS} collaboration, M.~Aaboud et~al., \emph{{Measurement of
  differential cross-sections of a single top quark produced in association
  with a $W$ boson at $\sqrt{s}=13$ TeV with ATLAS}},
  \href{https://doi.org/10.1140/epjc/s10052-018-5649-8}{\emph{Eur. Phys. J.}
  {\bfseries C78} (2018) 186},
  [\href{https://arxiv.org/abs/1712.01602}{{\ttfamily 1712.01602}}].

\bibitem{ATLAS:2020cwj}
{\scshape ATLAS} collaboration, G.~Aad et~al., \emph{{Measurement of single
  top-quark production in association with a $W$ boson in the single-lepton
  channel at $\sqrt{s} = 8\,\text {TeV}$ with the ATLAS detector}},
  \href{https://doi.org/10.1140/epjc/s10052-021-09371-7}{\emph{Eur. Phys. J. C}
  {\bfseries 81} (2021) 720},
  [\href{https://arxiv.org/abs/2007.01554}{{\ttfamily 2007.01554}}].

\bibitem{Chatrchyan:2012zca}
{\scshape CMS} collaboration, S.~Chatrchyan et~al., \emph{{Evidence for
  associated production of a single top quark and W boson in $pp$ collisions at
  $\sqrt{s}$ = 7 TeV}},
  \href{https://doi.org/10.1103/PhysRevLett.110.022003}{\emph{Phys. Rev. Lett.}
  {\bfseries 110} (2013) 022003},
  [\href{https://arxiv.org/abs/1209.3489}{{\ttfamily 1209.3489}}].

\bibitem{Chatrchyan:2014tua}
{\scshape CMS} collaboration, S.~Chatrchyan et~al., \emph{{Observation of the
  associated production of a single top quark and a $W$ boson in $pp$
  collisions at $\sqrt s = $8 TeV}},
  \href{https://doi.org/10.1103/PhysRevLett.112.231802}{\emph{Phys. Rev. Lett.}
  {\bfseries 112} (2014) 231802},
  [\href{https://arxiv.org/abs/1401.2942}{{\ttfamily 1401.2942}}].

\bibitem{Sirunyan:2018lcp}
{\scshape CMS} collaboration, A.~M. Sirunyan et~al., \emph{{Measurement of the
  production cross section for single top quarks in association with W bosons
  in proton-proton collisions at $ \sqrt{s}=13 $ TeV}},
  \href{https://doi.org/10.1007/JHEP10(2018)117}{\emph{JHEP} {\bfseries 10}
  (2018) 117}, [\href{https://arxiv.org/abs/1805.07399}{{\ttfamily
  1805.07399}}].

\bibitem{CMS:2021vqm}
{\scshape CMS} collaboration, A.~Tumasyan et~al., \emph{{Observation of tW
  production in the single-lepton channel in pp collisions at $ \sqrt{s} $ = 13
  TeV}}, \href{https://doi.org/10.1007/JHEP11(2021)111}{\emph{JHEP} {\bfseries
  11} (2021) 111}, [\href{https://arxiv.org/abs/2109.01706}{{\ttfamily
  2109.01706}}].

\bibitem{Rodriguez:2021ckm}
A.~S. Rodr\'\i{}guez, \emph{{Inclusive and differential cross-sections
  measurements in the single top tW e-mu channel with CMS}},
  \href{https://doi.org/10.22323/1.398.0444}{\emph{PoS} {\bfseries EPS-HEP2021}
  (2022) 444}, [\href{https://arxiv.org/abs/2112.08976}{{\ttfamily
  2112.08976}}].

\bibitem{Giele:1995kr}
W.~T. Giele, S.~Keller and E.~Laenen, \emph{{QCD corrections to $W$ boson plus
  heavy quark production at the Tevatron}},
  \href{https://doi.org/10.1016/0370-2693(96)00078-0}{\emph{Phys. Lett.}
  {\bfseries B372} (1996) 141--149},
  [\href{https://arxiv.org/abs/hep-ph/9511449}{{\ttfamily hep-ph/9511449}}].

\bibitem{Zhu:2001hw}
S.~Zhu, \emph{{Next-to-leading order QCD corrections to bg $\to$ tW${}^-$ at
  CERN large hadron collider}},
  \href{https://doi.org/10.1016/S0370-2693(02)01952-4,
  10.1016/S0370-2693(01)01404-6}{\emph{Phys. Lett.} {\bfseries B524} (2002)
  283--288}, [\href{https://arxiv.org/abs/hep-ph/0109269}{{\ttfamily
  hep-ph/0109269}}].

\bibitem{Cao:2008af}
Q.-H. Cao, \emph{{Demonstration of One Cutoff Phase Space Slicing Method:
  Next-to-Leading Order QCD Corrections to the tW Associated Production in
  Hadron Collision}},  \href{https://arxiv.org/abs/0801.1539}{{\ttfamily
  0801.1539}}.

\bibitem{Kant:2014oha}
P.~Kant, O.~M. Kind, T.~Kintscher, T.~Lohse, T.~Martini, S.~Mölbitz et~al.,
  \emph{{HatHor for single top-quark production: Updated predictions and
  uncertainty estimates for single top-quark production in hadronic
  collisions}}, \href{https://doi.org/10.1016/j.cpc.2015.02.001}{\emph{Comput.
  Phys. Commun.} {\bfseries 191} (2015) 74--89},
  [\href{https://arxiv.org/abs/1406.4403}{{\ttfamily 1406.4403}}].

\bibitem{Campbell:2005bb}
J.~M. Campbell and F.~Tramontano, \emph{{Next-to-leading order corrections to
  Wt production and decay}},
  \href{https://doi.org/10.1016/j.nuclphysb.2005.08.015}{\emph{Nucl. Phys.}
  {\bfseries B726} (2005) 109--130},
  [\href{https://arxiv.org/abs/hep-ph/0506289}{{\ttfamily hep-ph/0506289}}].

\bibitem{Kidonakis:2006bu}
N.~Kidonakis, \emph{{Single top production at the Tevatron: Threshold
  resummation and finite-order soft gluon corrections}},
  \href{https://doi.org/10.1103/PhysRevD.74.114012}{\emph{Phys. Rev.}
  {\bfseries D74} (2006) 114012},
  [\href{https://arxiv.org/abs/hep-ph/0609287}{{\ttfamily hep-ph/0609287}}].

\bibitem{Kidonakis:2010ux}
N.~Kidonakis, \emph{{Two-loop soft anomalous dimensions for single top quark
  associated production with a $W^-$ or $H^-$}},
  \href{https://doi.org/10.1103/PhysRevD.82.054018}{\emph{Phys. Rev.}
  {\bfseries D82} (2010) 054018},
  [\href{https://arxiv.org/abs/1005.4451}{{\ttfamily 1005.4451}}].

\bibitem{Kidonakis:2016sjf}
N.~Kidonakis, \emph{{Soft-gluon corrections for $tW$ production at N$^3$LO}},
  \href{https://doi.org/10.1103/PhysRevD.96.034014}{\emph{Phys. Rev.}
  {\bfseries D96} (2017) 034014},
  [\href{https://arxiv.org/abs/1612.06426}{{\ttfamily 1612.06426}}].

\bibitem{Kidonakis:2021vob}
N.~Kidonakis and N.~Yamanaka, \emph{{Higher-order corrections for $tW$
  production at high-energy hadron colliders}},
  \href{https://doi.org/10.1007/JHEP05(2021)278}{\emph{JHEP} {\bfseries 05}
  (2021) 278}, [\href{https://arxiv.org/abs/2102.11300}{{\ttfamily
  2102.11300}}].

\bibitem{Li:2019dhg}
C.~S. Li, H.~T. Li, D.~Y. Shao and J.~Wang, \emph{{Momentum-space threshold
  resummation in $tW$ production at the LHC}},
  \href{https://doi.org/10.1007/JHEP06(2019)125}{\emph{JHEP} {\bfseries 06}
  (2019) 125}, [\href{https://arxiv.org/abs/1903.01646}{{\ttfamily
  1903.01646}}].

\bibitem{Frixione:2008yi}
S.~Frixione, E.~Laenen, P.~Motylinski, B.~R. Webber and C.~D. White,
  \emph{{Single-top hadroproduction in association with a W boson}},
  \href{https://doi.org/10.1088/1126-6708/2008/07/029}{\emph{JHEP} {\bfseries
  07} (2008) 029}, [\href{https://arxiv.org/abs/0805.3067}{{\ttfamily
  0805.3067}}].

\bibitem{Re:2010bp}
E.~Re, \emph{{Single-top Wt-channel production matched with parton showers
  using the POWHEG method}},
  \href{https://doi.org/10.1140/epjc/s10052-011-1547-z}{\emph{Eur. Phys. J.}
  {\bfseries C71} (2011) 1547},
  [\href{https://arxiv.org/abs/1009.2450}{{\ttfamily 1009.2450}}].

\bibitem{Jezo:2016ujg}
T.~Ježo, J.~M. Lindert, P.~Nason, C.~Oleari and S.~Pozzorini, \emph{{An NLO+PS
  generator for $t\bar{t}$ and $Wt$ production and decay including non-resonant
  and interference effects}},
  \href{https://doi.org/10.1140/epjc/s10052-016-4538-2}{\emph{Eur. Phys. J.}
  {\bfseries C76} (2016) 691},
  [\href{https://arxiv.org/abs/1607.04538}{{\ttfamily 1607.04538}}].

\bibitem{Chen:2021gjv}
L.-B. Chen and J.~Wang, \emph{{Analytic two-loop master integrals for tW
  production at hadron colliders: I *}},
  \href{https://doi.org/10.1088/1674-1137/ac2a1e}{\emph{Chin. Phys. C}
  {\bfseries 45} (2021) 123106},
  [\href{https://arxiv.org/abs/2106.12093}{{\ttfamily 2106.12093}}].

\bibitem{Long:2021vse}
M.-M. Long, R.-Y. Zhang, W.-G. Ma, Y.~Jiang, L.~Han, Z.~Li et~al.,
  \emph{{Two-loop master integrals for the single top production associated
  with $W$ boson}},  \href{https://arxiv.org/abs/2111.14172}{{\ttfamily
  2111.14172}}.

\bibitem{Catani:2007vq}
S.~Catani and M.~Grazzini, \emph{{An NNLO subtraction formalism in hadron
  collisions and its application to Higgs boson production at the LHC}},
  \href{https://doi.org/10.1103/PhysRevLett.98.222002}{\emph{Phys. Rev. Lett.}
  {\bfseries 98} (2007) 222002},
  [\href{https://arxiv.org/abs/hep-ph/0703012}{{\ttfamily hep-ph/0703012}}].

\bibitem{Stewart:2010tn}
I.~W. Stewart, F.~J. Tackmann and W.~J. Waalewijn, \emph{{N-Jettiness: An
  Inclusive Event Shape to Veto Jets}},
  \href{https://doi.org/10.1103/PhysRevLett.105.092002}{\emph{Phys. Rev. Lett.}
  {\bfseries 105} (2010) 092002},
  [\href{https://arxiv.org/abs/1004.2489}{{\ttfamily 1004.2489}}].

\bibitem{Demartin:2016axk}
F.~Demartin, B.~Maier, F.~Maltoni, K.~Mawatari and M.~Zaro, \emph{{tWH
  associated production at the LHC}},
  \href{https://doi.org/10.1140/epjc/s10052-017-4601-7}{\emph{Eur. Phys. J.}
  {\bfseries C77} (2017) 34},
  [\href{https://arxiv.org/abs/1607.05862}{{\ttfamily 1607.05862}}].

\bibitem{Li:2016tvb}
H.~T. Li and J.~Wang, \emph{{Next-to-Next-to-Leading Order $N$-Jettiness Soft
  Function for One Massive Colored Particle Production at Hadron Colliders}},
  \href{https://doi.org/10.1007/JHEP02(2017)002}{\emph{JHEP} {\bfseries 02}
  (2017) 002}, [\href{https://arxiv.org/abs/1611.02749}{{\ttfamily
  1611.02749}}].

\bibitem{Li:2018tsq}
H.~T. Li and J.~Wang, \emph{{Next-to-next-to-leading order $N$-jettiness soft
  function for $tW$ production}},
  \href{https://doi.org/10.1016/j.physletb.2018.08.019}{\emph{Phys. Lett.}
  {\bfseries B784} (2018) 397--404},
  [\href{https://arxiv.org/abs/1804.06358}{{\ttfamily 1804.06358}}].

\bibitem{Hahn:2000kx}
T.~Hahn, \emph{{Generating Feynman diagrams and amplitudes with FeynArts 3}},
  \href{https://doi.org/10.1016/S0010-4655(01)00290-9}{\emph{Comput. Phys.
  Commun.} {\bfseries 140} (2001) 418--431},
  [\href{https://arxiv.org/abs/hep-ph/0012260}{{\ttfamily hep-ph/0012260}}].

\bibitem{Shtabovenko:2020gxv}
V.~Shtabovenko, R.~Mertig and F.~Orellana, \emph{{FeynCalc 9.3: New features
  and improvements}},
  \href{https://doi.org/10.1016/j.cpc.2020.107478}{\emph{Comput. Phys. Commun.}
  {\bfseries 256} (2020) 107478},
  [\href{https://arxiv.org/abs/2001.04407}{{\ttfamily 2001.04407}}].

\bibitem{Chen:2019wyb}
L.~Chen, \emph{{A prescription for projectors to compute helicity amplitudes in
  D dimensions}},
  \href{https://doi.org/10.1140/epjc/s10052-021-09210-9}{\emph{Eur. Phys. J. C}
  {\bfseries 81} (2021) 417},
  [\href{https://arxiv.org/abs/1904.00705}{{\ttfamily 1904.00705}}].

\bibitem{Korner:1991sx}
J.~G. Korner, D.~Kreimer and K.~Schilcher, \emph{{A Practicable gamma(5) scheme
  in dimensional regularization}},
  \href{https://doi.org/10.1007/BF01559471}{\emph{Z. Phys. C} {\bfseries 54}
  (1992) 503--512}.

\bibitem{Passarino:1978jh}
G.~Passarino and M.~J.~G. Veltman, \emph{{One Loop Corrections for e+ e-
  Annihilation Into mu+ mu- in the Weinberg Model}},
  \href{https://doi.org/10.1016/0550-3213(79)90234-7}{\emph{Nucl. Phys. B}
  {\bfseries 160} (1979) 151--207}.

\bibitem{Ellis:2007qk}
R.~K. Ellis and G.~Zanderighi, \emph{{Scalar one-loop integrals for QCD}},
  \href{https://doi.org/10.1088/1126-6708/2008/02/002}{\emph{JHEP} {\bfseries
  02} (2008) 002}, [\href{https://arxiv.org/abs/0712.1851}{{\ttfamily
  0712.1851}}].

\bibitem{Patel:2015tea}
H.~H. Patel, \emph{{Package-X: A Mathematica package for the analytic
  calculation of one-loop integrals}},
  \href{https://doi.org/10.1016/j.cpc.2015.08.017}{\emph{Comput. Phys. Commun.}
  {\bfseries 197} (2015) 276--290},
  [\href{https://arxiv.org/abs/1503.01469}{{\ttfamily 1503.01469}}].

\bibitem{Liu:2022chg}
X.~Liu and Y.-Q. Ma, \emph{{AMFlow: a Mathematica package for Feynman integrals
  computation via Auxiliary Mass Flow}},
  \href{https://arxiv.org/abs/2201.11669}{{\ttfamily 2201.11669}}.

\bibitem{Lee:2012cn}
R.~N. Lee, \emph{{Presenting LiteRed: a tool for the Loop InTEgrals
  REDuction}},  \href{https://arxiv.org/abs/1212.2685}{{\ttfamily 1212.2685}}.

\bibitem{Kotikov:1990kg}
A.~V. Kotikov, \emph{{Differential equations method: New technique for massive
  Feynman diagrams calculation}},
  \href{https://doi.org/10.1016/0370-2693(91)90413-K}{\emph{Phys. Lett.}
  {\bfseries B254} (1991) 158--164}.

\bibitem{Kotikov:1991pm}
A.~V. Kotikov, \emph{{Differential equation method: The Calculation of N point
  Feynman diagrams}}, \href{https://doi.org/10.1016/0370-2693(91)90536-Y,
  10.1016/0370-2693(92)91582-T}{\emph{Phys. Lett.} {\bfseries B267} (1991)
  123--127}.

\bibitem{Henn:2013pwa}
J.~M. Henn, \emph{{Multiloop integrals in dimensional regularization made
  simple}}, \href{https://doi.org/10.1103/PhysRevLett.110.251601}{\emph{Phys.
  Rev. Lett.} {\bfseries 110} (2013) 251601},
  [\href{https://arxiv.org/abs/1304.1806}{{\ttfamily 1304.1806}}].

\bibitem{Goncharov:1998kja}
A.~B. Goncharov, \emph{{Multiple polylogarithms, cyclotomy and modular
  complexes}}, \href{https://doi.org/10.4310/MRL.1998.v5.n4.a7}{\emph{Math.
  Res. Lett.} {\bfseries 5} (1998) 497--516},
  [\href{https://arxiv.org/abs/1105.2076}{{\ttfamily 1105.2076}}].

\bibitem{Becher:2009cu}
T.~Becher and M.~Neubert, \emph{{Infrared singularities of scattering
  amplitudes in perturbative QCD}},
  \href{https://doi.org/10.1103/PhysRevLett.102.162001}{\emph{Phys. Rev. Lett.}
  {\bfseries 102} (2009) 162001},
  [\href{https://arxiv.org/abs/0901.0722}{{\ttfamily 0901.0722}}].

\bibitem{Becher:2009qa}
T.~Becher and M.~Neubert, \emph{{On the Structure of Infrared Singularities of
  Gauge-Theory Amplitudes}},
  \href{https://doi.org/10.1088/1126-6708/2009/06/081}{\emph{JHEP} {\bfseries
  06} (2009) 081}, [\href{https://arxiv.org/abs/0903.1126}{{\ttfamily
  0903.1126}}].

\bibitem{Becher:2009kw}
T.~Becher and M.~Neubert, \emph{{Infrared singularities of QCD amplitudes with
  massive partons}},
  \href{https://doi.org/10.1103/PhysRevD.79.125004}{\emph{Phys. Rev. D}
  {\bfseries 79} (2009) 125004},
  [\href{https://arxiv.org/abs/0904.1021}{{\ttfamily 0904.1021}}].

\bibitem{Ferroglia:2009ep}
A.~Ferroglia, M.~Neubert, B.~D. Pecjak and L.~L. Yang, \emph{{Two-loop
  divergences of scattering amplitudes with massive partons}},
  \href{https://doi.org/10.1103/PhysRevLett.103.201601}{\emph{Phys. Rev. Lett.}
  {\bfseries 103} (2009) 201601},
  [\href{https://arxiv.org/abs/0907.4791}{{\ttfamily 0907.4791}}].

\bibitem{Mitov:2010xw}
A.~Mitov, G.~F. Sterman and I.~Sung, \emph{{Computation of the Soft Anomalous
  Dimension Matrix in Coordinate Space}},
  \href{https://doi.org/10.1103/PhysRevD.82.034020}{\emph{Phys. Rev. D}
  {\bfseries 82} (2010) 034020},
  [\href{https://arxiv.org/abs/1005.4646}{{\ttfamily 1005.4646}}].

\bibitem{Kidonakis:2019nqa}
N.~Kidonakis, \emph{{Soft anomalous dimensions for single-top production at
  three loops}}, \href{https://doi.org/10.1103/PhysRevD.99.074024}{\emph{Phys.
  Rev. D} {\bfseries 99} (2019) 074024},
  [\href{https://arxiv.org/abs/1901.09928}{{\ttfamily 1901.09928}}].

\bibitem{Li:2013mia}
H.~T. Li, C.~S. Li, D.~Y. Shao, L.~L. Yang and H.~X. Zhu, \emph{{Top quark pair
  production at small transverse momentum in hadronic collisions}},
  \href{https://doi.org/10.1103/PhysRevD.88.074004}{\emph{Phys. Rev. D}
  {\bfseries 88} (2013) 074004},
  [\href{https://arxiv.org/abs/1307.2464}{{\ttfamily 1307.2464}}].

\bibitem{Xu:2018eos}
X.~Xu and L.~L. Yang, \emph{{Towards a new approximation for pair-production
  and associated-production of the Higgs boson}},
  \href{https://doi.org/10.1007/JHEP01(2019)211}{\emph{JHEP} {\bfseries 01}
  (2019) 211}, [\href{https://arxiv.org/abs/1810.12002}{{\ttfamily
  1810.12002}}].

\bibitem{Wang:2020nnr}
G.~Wang, Y.~Wang, X.~Xu, Y.~Xu and L.~L. Yang, \emph{{Efficient computation of
  two-loop amplitudes for Higgs boson pair production}},
  \href{https://doi.org/10.1103/PhysRevD.104.L051901}{\emph{Phys. Rev. D}
  {\bfseries 104} (2021) L051901},
  [\href{https://arxiv.org/abs/2010.15649}{{\ttfamily 2010.15649}}].

\bibitem{Cornwall:1974km}
J.~M. Cornwall, D.~N. Levin and G.~Tiktopoulos, \emph{{Derivation of Gauge
  Invariance from High-Energy Unitarity Bounds on the s Matrix}},
  \href{https://doi.org/10.1103/PhysRevD.10.1145}{\emph{Phys. Rev. D}
  {\bfseries 10} (1974) 1145}.

\bibitem{Lee:1977eg}
B.~W. Lee, C.~Quigg and H.~B. Thacker, \emph{{Weak Interactions at Very
  High-Energies: The Role of the Higgs Boson Mass}},
  \href{https://doi.org/10.1103/PhysRevD.16.1519}{\emph{Phys. Rev. D}
  {\bfseries 16} (1977) 1519}.

\bibitem{Caswell:1974gg}
W.~E. Caswell, \emph{{Asymptotic Behavior of Nonabelian Gauge Theories to Two
  Loop Order}}, \href{https://doi.org/10.1103/PhysRevLett.33.244}{\emph{Phys.
  Rev. Lett.} {\bfseries 33} (1974) 244}.

\bibitem{Jones:1974mm}
D.~R.~T. Jones, \emph{{Two Loop Diagrams in Yang-Mills Theory}},
  \href{https://doi.org/10.1016/0550-3213(74)90093-5}{\emph{Nucl. Phys. B}
  {\bfseries 75} (1974) 531}.

\bibitem{Broadhurst:1991fy}
D.~J. Broadhurst, N.~Gray and K.~Schilcher, \emph{{Gauge invariant on-shell
  Z(2) in QED, QCD and the effective field theory of a static quark}},
  \href{https://doi.org/10.1007/BF01412333}{\emph{Z. Phys. C} {\bfseries 52}
  (1991) 111--122}.

\bibitem{Czakon:2007wk}
M.~Czakon, A.~Mitov and S.~Moch, \emph{{Heavy-quark production in gluon fusion
  at two loops in QCD}},
  \href{https://doi.org/10.1016/j.nuclphysb.2008.02.001}{\emph{Nucl. Phys. B}
  {\bfseries 798} (2008) 210--250},
  [\href{https://arxiv.org/abs/0707.4139}{{\ttfamily 0707.4139}}].

\bibitem{Czakon:2007ej}
M.~Czakon, A.~Mitov and S.~Moch, \emph{{Heavy-quark production in massless
  quark scattering at two loops in QCD}},
  \href{https://doi.org/10.1016/j.physletb.2007.06.020}{\emph{Phys. Lett. B}
  {\bfseries 651} (2007) 147--159},
  [\href{https://arxiv.org/abs/0705.1975}{{\ttfamily 0705.1975}}].

\bibitem{Melnikov:2000zc}
K.~Melnikov and T.~van Ritbergen, \emph{{The Three loop on-shell
  renormalization of QCD and QED}},
  \href{https://doi.org/10.1016/S0550-3213(00)00526-5}{\emph{Nucl. Phys. B}
  {\bfseries 591} (2000) 515--546},
  [\href{https://arxiv.org/abs/hep-ph/0005131}{{\ttfamily hep-ph/0005131}}].

\end{thebibliography}\endgroup
\bibliographystyle{JHEP}

\end{document}